# Water Entry Dynamics of Superhydrophobic and Lubricant-Impregnated Surfaces: A Theoretical and Experimental Perspective

Abhishek Mund, Shubham S. Ganar, Arindam Das[*]

School of Mechanical Sciences, Indian Institute of Technology (IIT) Goa, GEC Campus, Farmagudi, Ponda, Goa, 403401, India

*Email: arindam@iitgoa.ac.in

## Abstract

In this current work, a comprehensive investigation into the impact dynamics of superhydrophobic and lubricant-impregnated aluminium sphere balls on liquid surfaces were carried out. The study delves into the hydrodynamic lubrication behavior and the resulting phenomena during impact. The Worthington jet and cavity associated give more insight into fundamental forces. The experimental analysis and theoretical modeling suggest potential applications in various fields such as fluid dynamics, tribology, and materials science. The stability investigation on an aluminum-textured sphere ball is carried out, and the evaluation of jet height and pinch-off phenomena is analyzed in detail. The experiment on entrapped air on a superhydrophobic surface and the oil inside the LIS for greater height was performed. The experimental setup involved a carefully designed apparatus for impact analysis by utilizing high-speed imaging techniques. Parameters such as impact velocity, Weber number, and sphere diameter were systematically varied to understand their influence on the observed dynamics. The lubricant viscosity is chosen such that it is comparable to water viscosity. Furthermore, the study on Lubricant Impregnated Surfaces (LIS) assessed frictional losses, i.e., drag reduction, compared to control and textured surfaces. In conclusion, the valuable insights into the impact dynamics advance our understanding of hydrodynamic lubrication and its applications. The findings presented here underscore the importance of further research in optimizing surface properties, selecting suitable lubricants, and exploring broader applications in the fields of fluid dynamics, surface engineering, and mechanical engineering.

## 1. Introduction

Worthington and Cole[1] first studied the water entry of solids. The use of instantaneous photography in this context to determine the cavity and splash resulting from the impact of spheres on liquid surfaces in the latter part of the nineteenth century. Further investigations were conducted, and numerous researchers and engineers produced significant contributions to theoretical modelling. It has a broad spectrum of applications, including impact on a seaplane landing on water[2–4], defense[5,6], impact loading of offshore structures[7,8], water entry on missile projectiles[9], and so on. Several bio-inspired studies, such as seabirds executing plunge dives without injury[10,11], foxes diving into snow[12], and the slamming dynamics of diving fronts highlight a unified understanding of slamming forces across both animal and human systems[13]. In nature, water entry and exit are commonly observed in aquatic animals. Some leap out to catch their food source, while others dive beneath the surface to hunt. These dynamic air-water interactions have significant engineering implications, particularly in naval architecture, submarine design, and the development of ships and amphibious vehicles[14].

Water entry problems are essentially categorized as impact loading and cavity formation. The present research investigates the formation of cavities on various engineered surfaces. In recent years, improved photography/measurement tools have aided in the analysis of physical insights into water entry problems. Water entry on various shapes of objects[15,16], surface modifications[17], liquid pool conditions[18,19], hollow cylinder[20], and reduced drag on underwater traveling[21–25]. A few investigations on cavity formation at low bond number[26], and vertical water entry on low Froude number[27] have also been reported. At low Weber number and Reynolds number, it is found that surface conditions such as equilibrium contact angle have an influence on cavity formation. It is obvious that the capillary and viscous forces play a vital role[17]. Pinch-off phenomena[28] are observed when the air cavity interface collapses to a single point as a result of hydrostatic pressure; pinch-off depth can be dependent on the sphere's density. The higher the density, the higher the pinch-off depth, invariant with the pinch-off time.

Conventional surfaces pose challenges in solid-impact-on-liquid studies due to the air collapsing under high-speed impacts, leading to wetting transitions and loss of repellency. The trapped air layer (plastron) introduces unpredictability in cavity formation, splash dynamics, and Worthington jet behavior, complicating reproducibility. More durable alternatives, such as LIS, may provide better control for such studies. LIS has gained widespread acceptance in recent years because of its distinct features and applications. These surfaces have been engineered to hold lubricant inside their micro/nanostructures, enabling self-replenishment and long-term low-friction conditions even under demanding operating conditions[29–32]. The theoretical basis of non-wetting surfaces, such as superhydrophobic surfaces, originates from natural systems, like the lotus leaf, which has exceptional water-repelling properties due to its micro- and nanoscale surface patterns. Among the many different combinations of LIS, lubricant-impregnated aluminum sphere balls stand out as a promising possibility for enhancing lubrication performance. Aluminium, with its outstanding mechanical properties and vast industrial applications, is an excellent substrate for lubricant impregnation.

The current study investigates how pinch-off phenomena and cavities take place when different diameter sphere balls gently fall from different heights onto the water surface. How surface energy modification and roughness affect water entry dynamics. According to some prior investigations, cavity and pinch-off phenomena are distinct between roughness and smooth surfaces. However, introducing LIS to this phenomenon is novel, and it will be intriguing to observe its dynamics. The theoretical modeling incorporates the various types of forces that are involved in the dynamic process. Non-dimensional parameters are essential for understanding the dynamics of solid impact on liquid surfaces, as they help identify dominant physical mechanisms and simplify complex interactions. Key dimensionless numbers, such as the Weber number, Reynolds number, and Froude number, govern the balance between inertial, viscous, gravitational, and surface tension forces. By non-dimensionalizing governing equations and boundary conditions, the problem can be scaled to different geometries and fluid properties while preserving dynamic similarity. For instance, the Weber number characterizes the ratio of inertial to surface tension effects, while the Reynolds number compares inertial to viscous forces, influencing splash formation and cavity dynamics. This approach enables the generalization of experimental and numerical results across varying impact velocities, densities, and length scales, providing critical insights into phenomena like droplet splashing, crown formation, and jetting behavior in applications ranging from raindrop impacts to industrial spray processes.

The schematic of the fabricated experimental setup (see figure 2). A plexiglass (transparent) material tank is fabricated with the dimensions $400x400x1200\ mm^3$ ($lxbxh$) with a thickness of 8 mm. The spherical objects were made hydrophobic by a chemical vapor deposition process (more in the experimental section). To make LIS, a thin oil layer was coated on the textured surface employing a dip coater. The contact angle is measured on these surfaces, followed by using in the experiment. The Weber number is varied from a very low value, i.e., 10 to 100. The spherical object density and diameter eventually varied; it helps in providing a parametric investigation on the same. Further analysis involves dynamics in the cavity and pinch-off formation. The article provides an overview of utilizing LIS in an impact study.

## 2. Theoretical analysis

Two pivotal studies have shaped the understanding of cavity formation and dynamics during water entry, effectively defining the regimes that categorize subsequent research in this area. The first, conducted by Duez et al.[17], significantly extended the earlier work of May[9] by clarifying the conditions under which impacting spheres generate air cavities when they do not. It demonstrated that smooth spheres only form cavities when their impact velocity exceeds a critical threshold, which is dependent on the (advancing) static contact angle. A few years later, Aristoff and Bush[33] provided further insights by categorizing four distinct cavity types that can develop when the conditions for cavity formation are satisfied. They investigated spheres with a consistent contact angle of approximately 120°, systematically varying both impact velocities and sphere sizes. Their comprehensive dataset revealed that cavities could form across all tested velocities, with four specific cavity shapes identified based on their collapse or pinch-off behavior. These cavity types were each found to occupy distinct regions within a Bond–Weber number plot. In summary, it identified the transition from non-cavity to deep-seal regimes and the key factors governing cavity depth and pinch-off. As mentioned earlier, the water entry of spheres is a classic fluid dynamics problem with applications in naval and biological systems. Initial studies documented cavity formation and splash behavior during impacts. Later studies explored the effects of sphere size, surface properties, impact velocity, and fluid characteristics. Truscott et al.[34] showed that spin notably affects cavity formation, splash asymmetry, and critical velocity thresholds. Gekle and Gordillo[35] used simulations to analyze cavity collapse and Worthington jet formation, emphasizing the roles of Froude number and surface tension. Shi et al. highlighted the influence of nose shape, impact angle, and velocity on cavity evolution and stability. Recent research has also focused on spheres with textured and lubricant-impregnated surfaces, revealing further complexity in water entry dynamics.

When an object falls in a liquid pool, it is initially accelerated because of the gravitational force acting on it. As it accelerated, thus, the drag force increased. However, as the velocity of an object increases, the drag force increases until it reaches a condition where it is equal in magnitude but opposite in direction to the gravitational force. At this condition, the net force on the object is zero, resulting in a constant velocity known as terminal velocity. Under these conditions, an equilibrium state exists among all the forces acting on the moving sphere. It can be expressed as

$$\sum F = F_G - F_B - F_D = 0 \qquad (1)$$

$$F_G = -\frac{\pi}{6}\rho_p d_p^3 g$$

$$F_B = +\frac{\pi}{6}\rho_f d_p^3 g$$

$$F_D = +\frac{\pi}{8}\rho_f V_p^2 d_p^2 C_D$$

$F_G$: Gravity force, $F_B$: Buoyant force, $F_D$: Fluid drag

$\rho_p$: solid sphere density, $\rho_f$: fluid density, $V_p$: solid sphere (terminal) velocity,

$d_p$: solid sphere diameter, $g$: acceleration due to gravity, $C_D$: drag coefficient

While the relationship between terminal velocity and drag force provides a useful framework for understanding the behavior of falling objects in fluids, it's important to consider these assumptions and limitations when applying the concept to real-world scenarios. There are some assumptions for the above-derived relationship[36], such as steady-state flow, Newtonian fluid, laminar flow, spherical object, and negligible buoyancy. In addition, there are some limitations, such as dependence on fluid properties, dependence on object properties, and complex flow conditions. The trajectory analysis provides more information about the terminal velocity. The deceleration or loss of energy can be expressed in a mathematical form[33], which includes gravity, capillary, buoyant, and hydrodynamic force balance with the mass of the sphere. The dimensional analysis[37] using the Buckingham pi theorem for the study of dynamics is dependent on the density ratio, the Froude number, and the Weber number. Investigation of cavity dynamics resulting from projectile water entry with varying physical parameters is also reported[38].

When an object falls through a fluid, it initially accelerates due to the gravitational force acting on it. As it accelerates, the drag force exerted by the fluid increases. However, as the object's velocity increases, so does the drag force until it reaches a point where the drag force becomes equal in magnitude but opposite in direction to the gravitational force. At this point, the net force on the object becomes zero, resulting in a constant velocity known as the terminal velocity[36,39]. In other words, the scalar momentum equation of a ball moving through a fluid, expressed along the direction of motion or pathline, represents a balance between the inertia force and the various forces acting on the ball. These include the body force (such as gravity), the drag force due to fluid resistance, the added mass, and the Boussinesq-Basset term. By solving this equation, and under the assumption that both the Boussinesq-Basset term and the added mass term are negligible, a simplified expression for the terminal velocity of the ball is obtained[40]. This expression describes the constant velocity attained by the ball when the net force acting on it becomes zero.

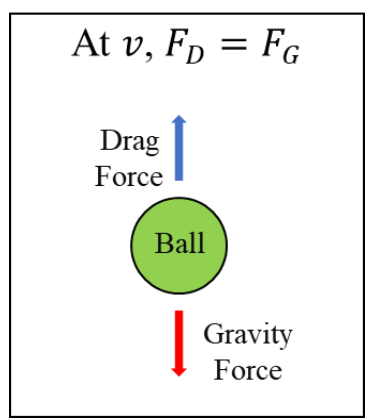

**Figure 1.** Schematics of terminal velocity

$$F_D = \frac{1}{2}\rho C_D v^2 A \quad (2)$$

$$F_G = mg \quad (3)$$

Where $C_D$: coefficient of drag, $A$: reference area, $g$: Acceleration due to gravity, $v$: velocity of the object relative to the fluid, $F_D$: drag force, $F_G$: gravitational force, $\rho$: density of the fluid, $m$: mass

At terminal velocity, the drag force acting on a falling object balances the gravitational force, resulting in a constant velocity. This analysis typically assumes steady-state flow, a Newtonian fluid, laminar flow conditions, negligible buoyancy effects, and a spherical object geometry. While these assumptions simplify the analysis and are valid for controlled scenarios, the model has limitations. Negligible buoyancy forces, which may become significant in certain fluid environments, can't accurately capture complex flow patterns and turbulence. Additionally, the terminal velocity is highly dependent on both object and fluid properties, such as density, viscosity, and surface roughness, which may limit the direct applicability of the simplified model in more complex or real-world systems.

Worthington and Cole[1] performed the water entry experiments with solid spheres and liquid droplets for the first time. Guleria et al.[37] studied water entry dynamics experiments on hydrophobic spheres and concluded their study with these key observations: the trajectory in the water after the water entry, followed by the formation of a trapped air cavity in the water pool. The subsequent cavity retraction and the emergence of Worthington jets, and the appearance of ripples and early bubble shedding from the bubbles attached to the descending spheres. These findings highlight the complex fluid dynamics involved during water entry events. Shi et al.[38] concluded that the cavity dynamics during water entry are significantly influenced by the nose shape, impact velocity, and impact angle of the impacting object. Their study systematically explored how these parameters affect the resulting cavity shapes and deformation patterns. Additionally, they characterized the cavity pinch-off process and the formation of cavity ripples that emerge following pinch-off, providing valuable insights into the transient fluid behavior and instabilities associated with high-speed impact events.

In summary, the emphasis is that the interaction between the lubricant and the surface microstructure, governed by surface tension, plays a critical role in influencing lubricant flow and retention, thereby affecting the stability and performance of the LIS under dynamic conditions. The external pressures from mechanical contact or fluid flow cause lubricant drainage and redistribution. Here, the research gap is using the LIS surface for a similar experiment set, in addition to stability analysis using the wettability properties.

**Table 1.** Phenomenon with key results and corresponding references

| **Key results** | **References** |
|---|---|
| Water entry experiments were conducted for the first time using both solid spheres and liquid droplets, focusing on key phenomena such as cavity formation, Worthington jet generation, and cavity surface ripples following closure. | Worthington and Cole[1] |
| The influence of contact angle and impact velocity was systematically examined to assess the water-repellent behavior of the spheres. | Duez et.al.[17] |
| Experimentally alter the Bond number, Weber number, and contact angle of smooth spheres; two key results: first, cavity shape also depends on the contact angle; second, the absence of a splash crown at low Weber number results in cavity formation below the predicted critical velocity. | Truscott et.al.[34] |
| Particular emphasis was to characterize the shape of the resulting air cavity in the low Bond number limit, where cavity collapse is driven principally by surface tension rather than gravity. | Aristoff and Bush[33] |
| A detailed investigation of Worthington jet formation was performed through boundary integral simulations, potential flow theory, and experimental validation. The jet base was identified as a critical zone for feeding the jet, and its role in jet dynamics was thoroughly discussed, covering axial acceleration, ballistic, and tip regions along the jet's development length. | Gekle and Gordillo[41] |
| Concluded four very important points, such as trajectory in the water after the water entry, trapped air-cavity formation within the water pool, cavity retraction, Worthington-jets formation, ripples, and early bubble shedding on the bubble attached to the descending spheres. | Guleria et.al.[37] |
| The effects of nose shape, impact velocity, and impact angle on cavity shapes are explored. Cavity pinch-off and the cavity ripples formed after pinch-off are characterized. | Shi et.al.[38] |
| The study introduced the analysis of free-body trajectories using a natural coordinate system, enabling independent, real-time measurement of lift and drag forces. It further proposed specific characteristics of time-dependent flow separation around the decelerating sphere, based on the observed force data. | Billa et.al.[40] |
| Early drag coefficient collapse on SLIPs, followed by a Hydrophobic surface. 30% higher Re required for hydrophobic for the same sharp fall compared to SLIPS. Lamella detachment condition and its behaviour, critical Reynolds number for cavity formation. | Zhu et.al.[23] |

## 3. Results and Discussions

### 3.1. Details of experimental setup and non-dimensional parameters

The Table 2. presents data related to a sphere's density and the liquid's density in which it is submerged. Specifically, it includes measurements for the diameter of an aluminum spherical ball and the height from which the ball is released. These parameters are critical in understanding the dynamics of the sphere's motion and its interaction with the liquid medium. The water entry experiments were conducted at a water temperature of approximately 20°C, with the fluid properties being: water density ($\rho_l$) of 1000 $kg/m^3$, the dynamic viscosity ($\mu_l$) of water, 1.002 $mPa.S$, and the surface tension is ($\gamma_l$) 72 $mN/m$. Based on the release height, the impact velocity of the ball is calculated. The impact velocity at the air–water interface is calculated using the equation $v = \sqrt{2gh}$ where v is the velocity of the sphere at the impact and h is the release height above the air–water interface. The release heights used in this reported work were 5, 20, and 40 cm. The spheres employed were made of aluminum, with a density ($\rho_s$) of 2700 $kg/m^3$. The aluminum sphere's density with water density has a density ratio of 2.70. Various parameters are analyzed as the object is released from distinct heights 5, 20, and 40 cm. The sphere diameters were varied at 4, 6, and 8 mm with a tolerance of $\pm 0.04 mm$.

The experimental setup consists of several components (Figure 2). The schematic of the fabricated experimental setup is shown below. The solid-impact experiment was carried out within a transparent plexiglass. A plexiglass (transparent) material tank is fabricated with the dimensions $400x400x1200$ $mm^3$ ($lxbxh$) with a thickness of 8 mm. The large size of the tank ensures that the water waves generated by the sphere's impact do not interfere with the observed phenomena. Additionally, the current experimental setup satisfies the condition related to the Froude number $Fr < 0.21(l/d)^2$, where d is the diameter of the sphere. The plexiglass was deeply cleaned with isopropyl alcohol solvents before being filled with water. One side of the spherical ball was attached to a solenoid valve, which was positioned atop a stand. A particular aluminium sphere ball is then dropped from a height with the help of a solenoid valve. A Phantom VEO 410 high-speed camera with a resolution of 1280 × 720 pixels and 5000 frames per second was employed to record the videos. The camera is connected to the computer via a LAN (Local Area Network) cable and is manually triggered using the PCC (Phantom Camera Control) software. Initial video analysis uses the PCC software, while object path tracking is performed using MATLAB. A high-beam light source was placed behind the substrate so that the light source, substrate, and high-speed camera were on the same optical axis.

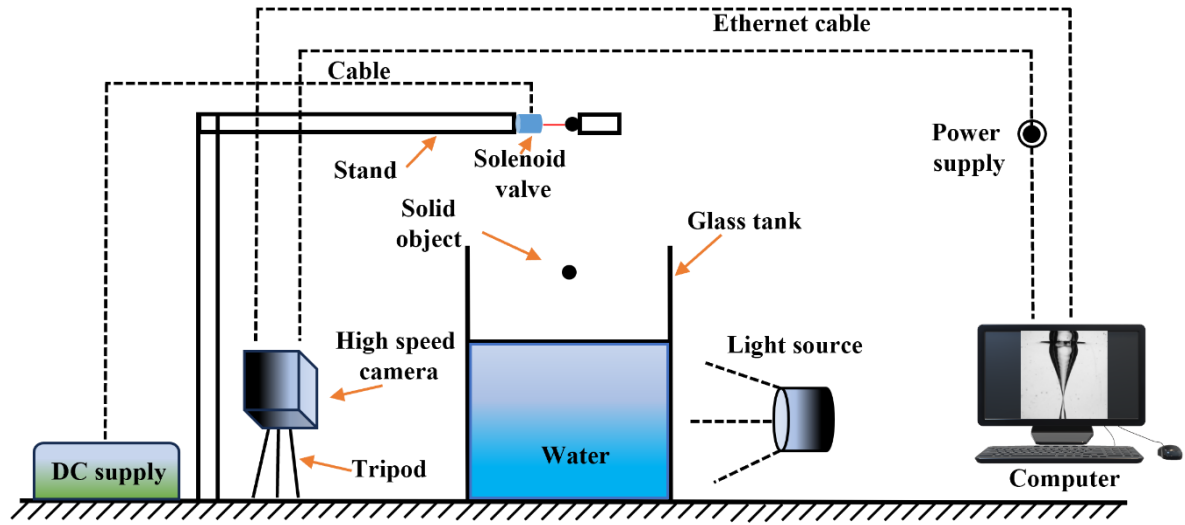

**Figure 2.** Schematics of experimental setup.

A summary of impact velocities along with key dimensionless parameters is provided in Table 2 below for a 4 mm diameter spherical object. As the release height increases, the impact velocity rises accordingly from 0.44 m/s to 2.8 m/s. This increase in velocity also results in higher dimensionless numbers (To determine the energy and momentum it carries upon contact with the liquid surface): Weber number (We), Froude number (Fr), and Reynolds number (Re), indicating a more dominant role of inertial forces over surface tension, gravity, and viscous forces, respectively. The Weber number, a dimensionless parameter that quantifies the relative importance of inertial forces to surface tension forces. This number plays a crucial role in impact studies as it helps analyze phenomena such as splashing, droplet formation, and deformation at the liquid interface. The Bond number (Bo), which remains constant at 2.17, signifies a fixed balance between gravitational and surface tension forces due to constant diameter and fluid properties. Similar observations were also made with spherical balls of other diameters, such as 6 and 8 mm.

**Table 2.** Properties and some non-dimensional parameters of different diameter spherical ball

| Density ratio ($\rho_s/\rho_l$) | Dia (mm) | Height (cm) | Velocity (m/s) | Bo | We | Fr | Re |
|---|---|---|---|---|---|---|---|
| 2.70 | 4 | 1 | 0.44 | 2.17 | 10.75 | 2.22 | 1973.96 |
| | | 5 | 0.99 | | 54.45 | 4.99 | 4441.42 |
| | | 20 | 1.98 | | 217.8 | 9.99 | 8882.85 |
| | | 40 | 2.8 | | 435.55 | 14.13 | 12561.61 |

| Density ratio ($\rho_s/\rho_l$) | Dia (mm) | Height (cm) | Velocity (m/s) | Bo | We | Fr | Re |
|---|---|---|---|---|---|---|---|
| 2.70 | 6 | 1 | 0.44 | 4.83 | 15.91 | 1.81 | 2960.95 |
| | | 5 | 0.99 | | 80.56 | 4.08 | 6662.14 |
| | | 20 | 1.98 | | 322.26 | 8.16 | 13324.28 |
| | | 40 | 2.8 | | 644.45 | 11.54 | 18842.42 |

| Density ratio ($\rho_s/\rho_l$) | Dia (mm) | Height (cm) | Velocity (m/s) | Bo | We | Fr | Re |
|---|---|---|---|---|---|---|---|
| 2.70 | 8 | 1 | 0.44 | 8.60 | 21.21 | 1.57 | 3947.93 |
| | | 5 | 0.99 | | 107.42 | 3.53 | 8882.85 |
| | | 20 | 1.98 | | 429.68 | 7.06 | 17765.71 |
| | | 40 | 2.8 | | 859.27 | 9.99 | 25123.23 |

## 3.2. Material Characterization

The pictures below show scanning electron microscope (SEM) micrographs of a micro- and nanostructured material at different magnifications. The image 3.a is a 2.5 M HCl etched sample, revealing a highly porous morphology, suggesting a high surface area architecture. The image 3.b is at a higher magnification with a 500 nm scale bar, highlighting nano-petal morphology. A conforming nano-petals structure is highly desirable to generate hierarchical morphologies suitable for applications, where the nanostructures may contribute to enhanced physical or chemical properties.

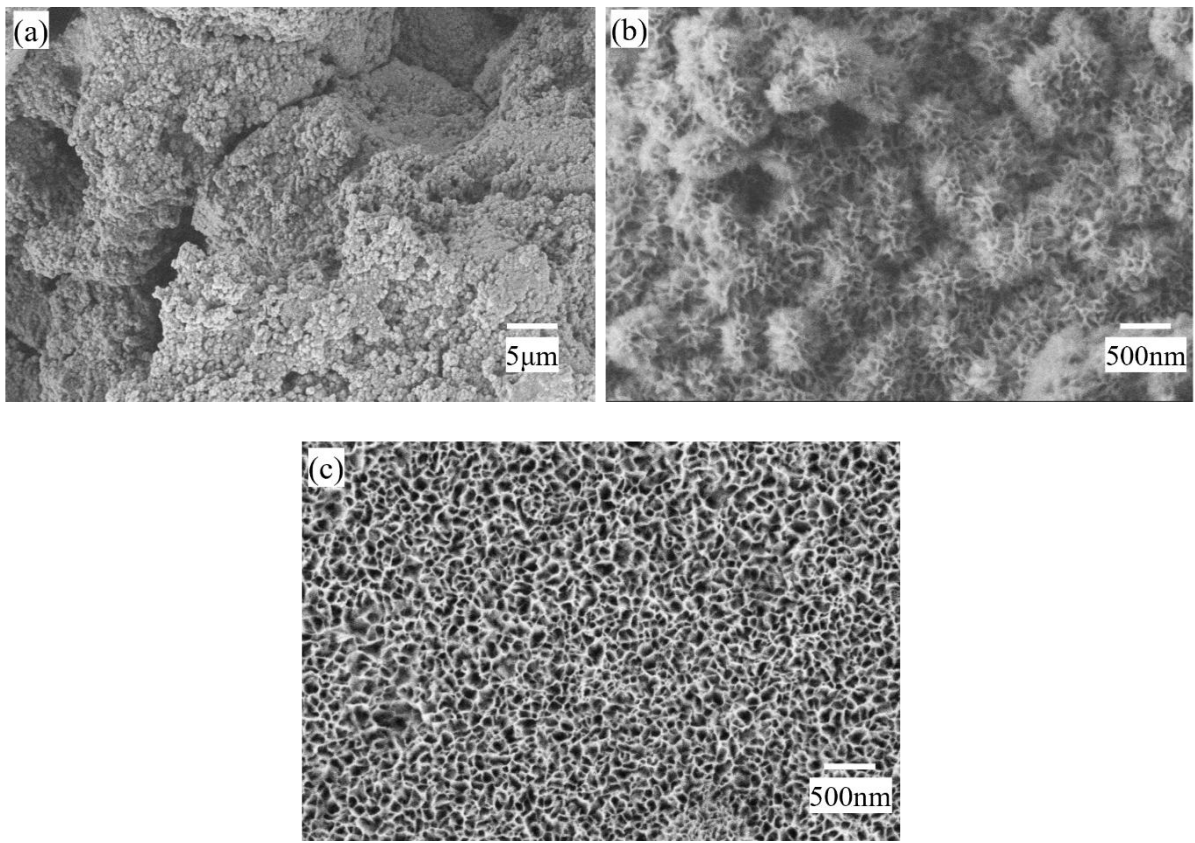

**Figure 3.** FESEM images showing the micro- and nanostructures of aluminum treated with (a) 2.5 M HCl and (c) 1 M HCl. Image (b) presents a magnified view of the sample etched with 2.5 M HCl.

The image 3.c, sample is a 1M HCl acid-etched sample, revealing the presence of densely packed nanosheets or nano petals morphology after boiling water treatment. It generates sparsely spaced micron-scale surface features.

The bar graph below (refer figure 4) illustrates the variation in water contact angles across three different surfaces: Control, Micro-nano LIS, and Micro-nano OTS. The control surface exhibits a moderate hydrophilicity with a contact angle of 70°, whereas the Micro-nano LIS surface shows an increased angle of 98°, indicating enhanced hydrophobicity due to the introduction of micro and nanostructures. Notably, the Micro-nano OTS surface achieves a significantly higher contact angle of 155°, demonstrating a transition to superhydrophobic behavior.

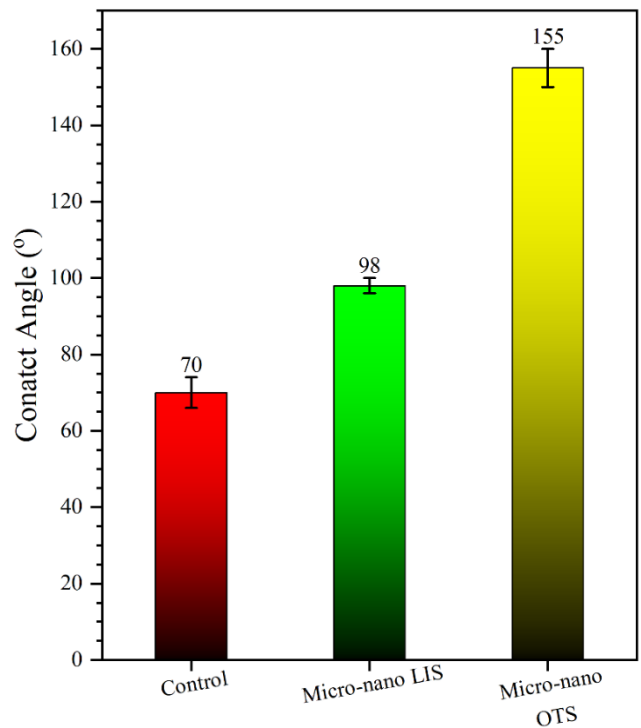

**Figure 4.** Wettability measurement on various samples.

## 3.3. Fabrication of LIS

Fabricated textured aluminium balls were impregnated with the lubricant (silicone oil) using a dip coating process. Dip coating velocities were carefully selected to ensure the absence of excess lubricant layers. The capillary number values less than $10^{-5}$ corresponds to the dip coating process ensuring the absence of excess lubricant layer[42] irrespective of surface roughness parameters. The dip coating velocity ($V$) was obtained using equation 4 and setting the value of the capillary number to $10^{-5}$. This capillary number's value can range from $10^{-5}$ to $10^{-4}$. In this case, we've evaluated $10^{-5}$.

$$\text{Capillary number (Ca)} = \frac{\mu V}{\gamma} = 10^{-4} \tag{4}$$

Where $\mu$= viscosity of the liquid, $\gamma$= surface tension, $V$= withdrawal velocity.

Table 3. Properties and Withdrawal velocity of lubricant liquid

| Lubricant | Liquid-Vapour Surface Tension $(mN/m)$ | Liquid Density $(kg/m^3)$ | Dynamic Viscosity $(mPa.s)$ | Withdrawal velocity $(mm/min)$ |
|---|---|---|---|---|
| **Silicone oil** | 19.7 | 913 | 5 | 25.8 |

As mentioned earlier, fabricating a stable LIS is the most important requirement for any application. This is characterized by one of the methods, namely the wettability measurement. The equilibrium contact angle of silicone oil on smooth aluminum in water must be measured and compared to the critical contact angle.

### 3.4. Evolution of various Cavity and Pinch off phenomena

The experiment was conducted using various surfaces. The observations were made at various heights to analyze the behavior of cavities formed by aluminum spherical balls of different diameters upon impact. The image (see figure 5) provides a comparative analysis of the impact behavior of a 6 mm diameter spherical aluminum ball, dropped from a height of

20 cm onto water, under varying surface conditions: wetting, functionalization, and LIS. The surfaces include a Control, Control with OTS, Nano-textured OTS, Micro-nano textured OTS, Nano-textured LIS, and Micro-nano textured LIS. The table below highlights the differences in surface roughness, functionalization, and lubricant presence across these samples. The key observations show significant variation in cavity formation: Control surfaces produce a quasi-static impact cavity, while OTS-functionalized and LIS surfaces generate a shallow impact cavity. Micronano and nano-functionalized surfaces produce a deep seal impact cavity. Notably, the LIS surfaces, which incorporate a lubricant layer, show minimized cavity formation and suppressed splash behavior. The results emphasize the strong influence of surface texture, chemical treatment, and lubrication on impact dynamics and cavity evolution. These phenomena can be observed as a result of the competing action of surface tension, which stretches the air-water contact and causes the ball to descend. In addition, it is visible that the air cavity formed on LIS is comparatively larger than the smooth one. Here, the ball diameter and releasing height are the same for all samples. Changing the height effectively changes the impact velocity, which affects pinch-off phenomena; also, ball diameter may play a role.

The formation of the Worthington jet above the water surface was observed across all samples (see supporting information). The jet dynamics were characterized by measuring the jet height from the water surface over time and correlating it with the impact velocity. The breakup of these jets may be attributed to Rayleigh-Plateau instability[43], which governs the disintegration of liquid columns due to surface tension. Among all the tested cases, the textured aluminum ball produced the highest jet height, while the smooth and LIS-coated samples showed significantly lower values. These differences in jet height can be associated with variations in interfacial forces at the air–water contact line, as well as differences in cavity collapse dynamics driven by hydrostatic pressure. Additionally, surface texture likely enhances upward momentum transfer during cavity collapse, leading to taller Worthington jets. The suppression of jet height in LIS surfaces suggests reduced energy focusing during the cavity rebound phase.

| | 6 mm diameter spherical aluminium ball dropped from a height of 20 cm | | | | | |
|---|---|---|---|---|---|---|
| **Sample** | **Control** | **Control + OTS** | **Nano-textured OTS** | **Micro-nano textured OTS** | **Nano-textured LIS** | **Micro-nano textured LIS** |
| | 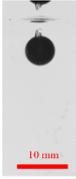 | 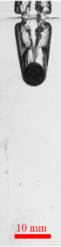 | 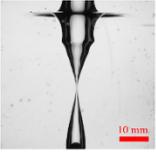 | 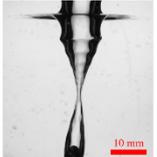 | 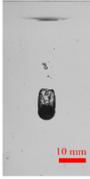 | 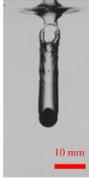 |
| **Roughness** | No | No | Nano scale | Micro-nano scale | Nano scale | Micro-nano scale |
| **Functionalization** | No | OTS | OTS | OTS | OTS | OTS |
| **Lubricant** | No | No | No | No | Silicone oil | Silicone oil |
| **Key observation** | Quasi-static cavity | Shallow seal cavity | Deep-seal cavity Standing capillary wave | Deep-seal cavity Standing capillary wave | Shallow seal cavity | Shallow seal cavity |

**Figure 5.** Effect of wetting, texture, and LIS on the cavity formation.

The difference in cavity formation highlights the role of surface functionalization in altering the solid-liquid interaction during impact. Such variations can significantly influence the dynamics of splash suppression and energy dissipation at the moment of impact.

### 3.4.1. Effect of wetting

The image (figure 6) sequence below illustrates the impact dynamics of spherical balls with varying diameters of 4, 6, and 8 mm released from 20 cm heights onto a liquid surface. For all sizes, the initial contact (at time *t*) triggers surface deformation and cavity formation. At the lowest height (see supporting information), all spheres show minimal splashing and gradually deform into spherical shapes with subtle surface waves. As the height increases to 20 cm, the impact velocity increases, leading to more pronounced deformation and the onset of air entrainment and bubble formation, especially for larger diameters. At 40 cm (see supporting information), larger jet height, and shallow cavity formation, particularly in the 6 mm and 8 mm diameter cases, indicating a strong dependence of impact dynamics on both sphere size and drop height. This progression emphasizes the interplay between kinetic energy, inertia, and surface tension in governing the outcome of solid impacts. The clear differences between the three rows highlight the scale dependence of splash dynamics and underscore the importance of diameter in determining impact outcomes on fluid surfaces.

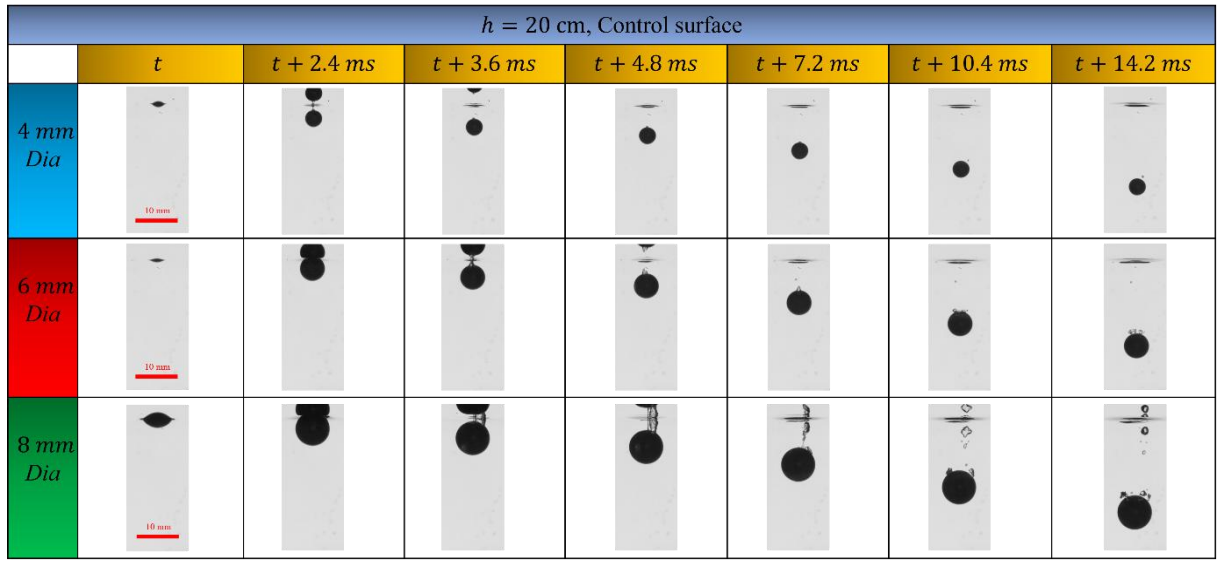

**Figure 6.** Quasi-static impact cavity. High-speed imaging of control aluminium sphere ball (4,6, and 8 mm dia) impact on liquid from various heights.

The image sequence below illustrates control surfaces functionalized with octadecyl trichlorosilane (OTS). Regardless of the diameter of the aluminum balls, the cavities formed by their impact were always small at a height of 5 cm (refer to supporting information Image S2.A). Interestingly, despite the small variance in cavity size, the pinch-off sites of the cavities were clearly evident, suggesting uniformity in the occurrence of these features. This cavity could possibly be referred to as a quasi-static impact cavity. There were notable variations when the height was raised to 20 cm (see figure 7). The cavities were significantly longer for aluminum spheres with 6 mm and 8 mm diameters, indicating that the impact energy at this height was adequate to lengthen the cavity. Shallow seal impact cavities are an acronym attributed to these kinds of cavities. However, the 4 mm diameter sphere's cavity maintained consistent with the 5 cm height, suggesting that even at a larger size, its lower mass constrained the cavity's expansion. At the maximum height of 40 cm (refer to supporting information Image S2.A), the cavities achieved their greatest lengths, showcasing a clear progression in cavity elongation with increasing height. The evolution of the cavities was particularly distinct at this height, with visible differences in the cavity formation and collapse dynamics. The cavities appear to be shallow seal impact cavities for all diameters. These observations underscore the influence of both sphere diameter and impact height on cavity behavior, highlighting how kinetic energy and mass interplay to produce varied outcomes.

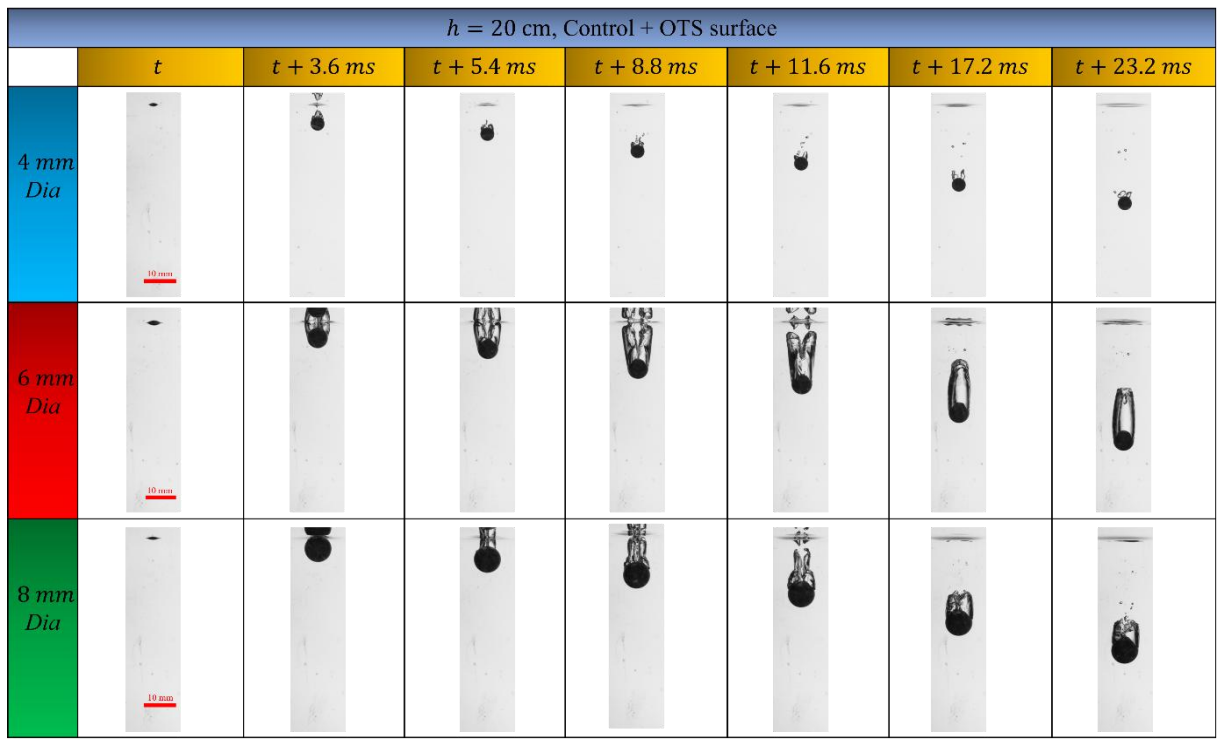

**Figure 7.** Quasi-static and shallow seal impact cavity. High-speed imaging of control OTS aluminium sphere ball (4,6, and 8 mm dia) impact on liquid from various heights.

### 3.4.2. Effect of texture

It is evident that the cavity shape changes for functionalized surfaces once texturing is introduced. The image below (refer figure 8) represents the air cavity profiles for micro-nano

OTS-treated surfaces impacted by 4-, 6-, and 8-mm diameter spheres dropped from a height of 20 cm, exhibiting distinct cavity evolution patterns influenced by sphere size and surface characteristics. For the 6- and 8-mm spheres, the cavity demonstrates significant radial expansion at the initial stages, followed by a gradual narrowing and elongation as time progresses toward pinch-off. The cavity profiles show minor asymmetries, suggesting the presence of instabilities. The formation of a wide splash crown indicates efficient momentum transfer and surface disruption due to the combined effects of micro- and nano-scale surface roughness with OTS functionalization. In contrast, the 8 mm sphere generates a deeper, more axisymmetric cavity with a higher degree of pinchoff depth. The cavity profiles remain relatively stable across successive time frames, indicating that the larger sphere's greater inertia helps maintain cavity symmetry and suppresses radial instabilities. The pinchoff was observed at 38.4 ms; however, sustained cavity walls were observed approximately 43.2 ms and beyond. It emphasizes the scale-dependent transition in cavity dynamics. Overall, the comparison (refer to supporting image S3.A) underscores that micro-nano structured hydrophobic surfaces can significantly modulate the air cavity morphology, with sphere size playing a critical role in governing pinchoff depth and radial spread.

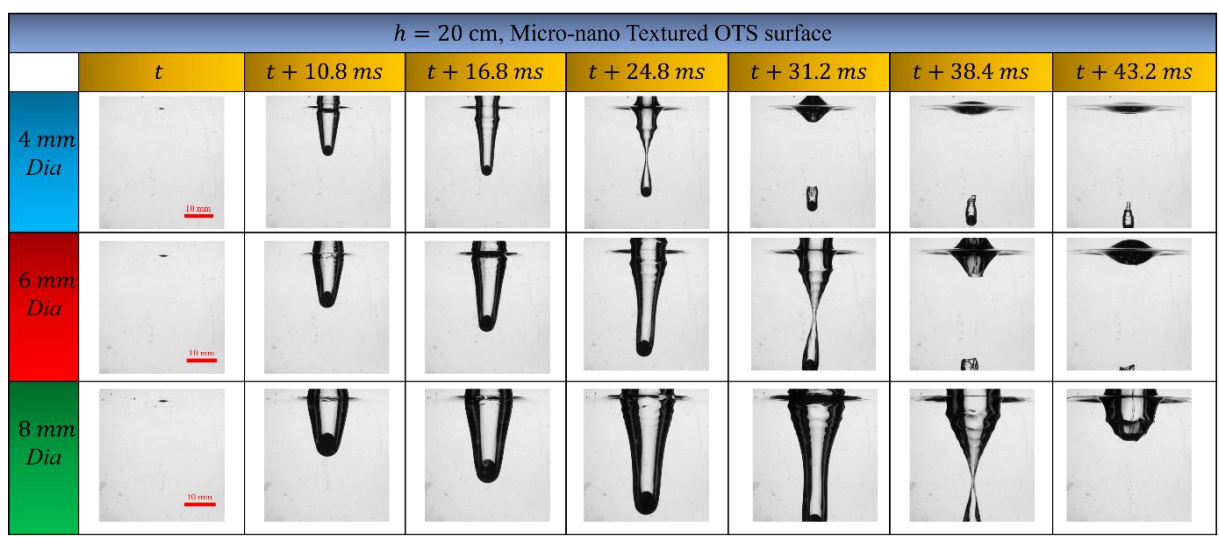

**Figure 8.** Deep-seal impact cavity. High-speed imaging of micro nano OTS aluminium sphere ball (4,6 and 8 mm dia) impact on liquid from various heights.

The plot below (see figure 9) shows the temporal evolution of cavity profiles. From the profile, it is clearly understood that the cavity initially deepens rapidly, reaching its maximum depth. Beyond this point, the cavity begins to retract and narrow, particularly near the bottom, suggesting the onset of cavity collapse. Lateral expansion at the upper section becomes more evident over time, as seen in the widening of the curves near the surface (z = 0 mm).

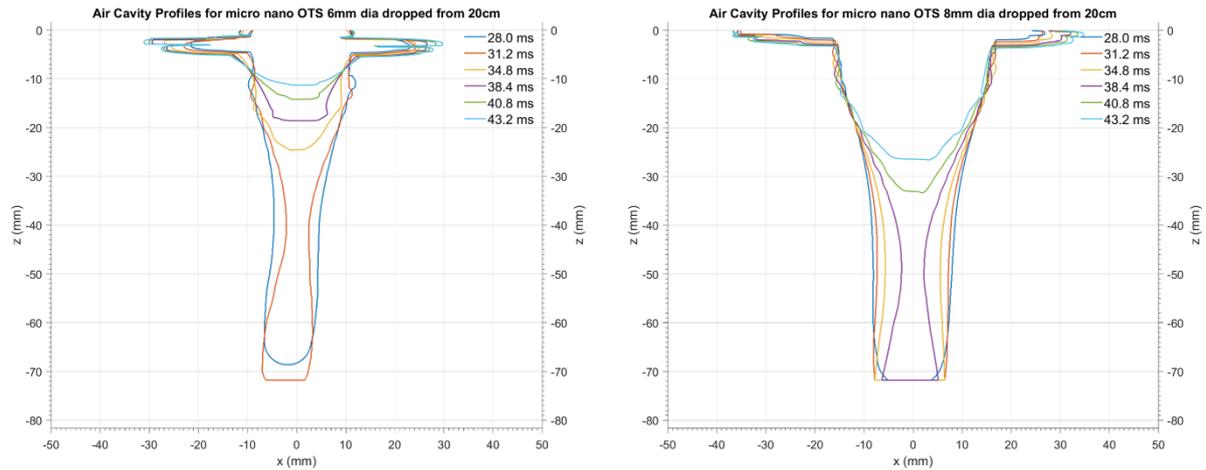

**Figure 9.** Temporal evolution of cavity profiles for micro-nano OTS dropped from 20 cm for the various sphere diameters.

Deep seal impact cavities are mostly noticeable on nano-textured OTS surfaces (see figure 10). At 5 cm high, a quasi-static impact cavity is observed for spherical balls with a diameter of 4 mm. Subsequently, a shallow seal impact cavity for spherical balls with dimensions of 6 and 8 mm. There is certainly evidence that the cavity's growth varies with its diameter. Only textured aluminum surfaces with functionalisation exhibit the deep seal cavity and associated pinch-off phenomena. Such a deep seal cavity and associated pinch-offs are not formed by the other set of samples, LIS and control. A strong capillary wave[44] phenomenon is observed during the water entry of cylindrical objects, attributed to the formation of standing capillary waves. These waves arise from the interaction of crests and troughs during the cavity contraction phase, as illustrated in the image (see supporting information Image S4.A, 8 mm diameter sphere dropped from a height of 5 cm). A similar observation was also made for the micro-nano OTS sample. The standing capillary waves result from the competing effects of surface tension, responsible for stretching the air-water interface and the descent of the sphere.

Owing to the unique dynamics involved in cavity formation, the influence of solid impact on liquids is of considerable interest. The fluid first expands radially before rapidly contracting due to fluid forces like surface tension and hydrostatic pressure; the timescale for all of these phenomena occurred in milliseconds. The time taken from cavity formation to its pinch-off can be categorized into three distinct timescales, each shedding light on the mechanisms underlying the observed pinch-off phenomena. Pinch-Off Time: The time it takes for the descending sphere to reach the pinch-off location. Cavity Expansion Timescale: This corresponds to the cavity's outward growth, driven by the kinetic energy of the descending sphere. Cavity Contraction Timescale: This is governed by the forces of surface tension or hydrostatic pressure, which act to close the cavity. These timescales provide a framework for understanding the primary factors influencing different types of pinch-off behavior.

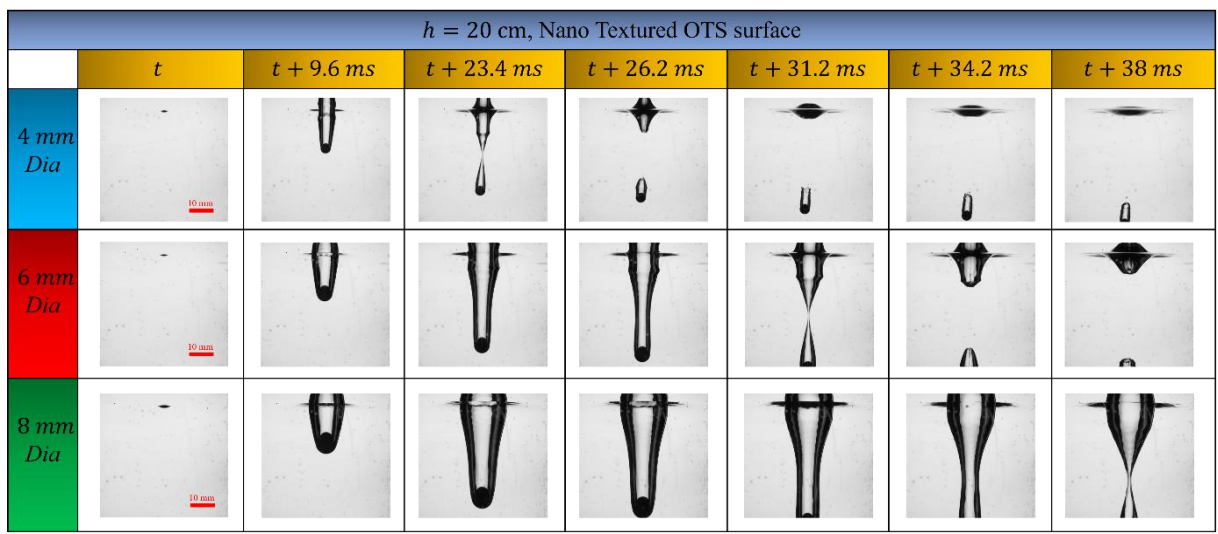

**Figure 10.** Deep seal impact cavity. High-speed imaging of nano-textured OTS aluminium sphere ball (4,6, and 8 mm dia) impact on liquid from various heights.

The temporal shift in cavity shapes between the two cases emphasizes the scale-dependent hydrodynamics, where the larger sphere exhibits more controlled cavity growth and collapse with reduced radial instability compared to the smaller sphere, likely due to the dominance of inertial forces over capillary effects (see figure 11). These observations highlight the influence of impact or size on air cavity dynamics, cavity stability, and pinch-off behavior on superhydrophobic nano-structured surfaces.

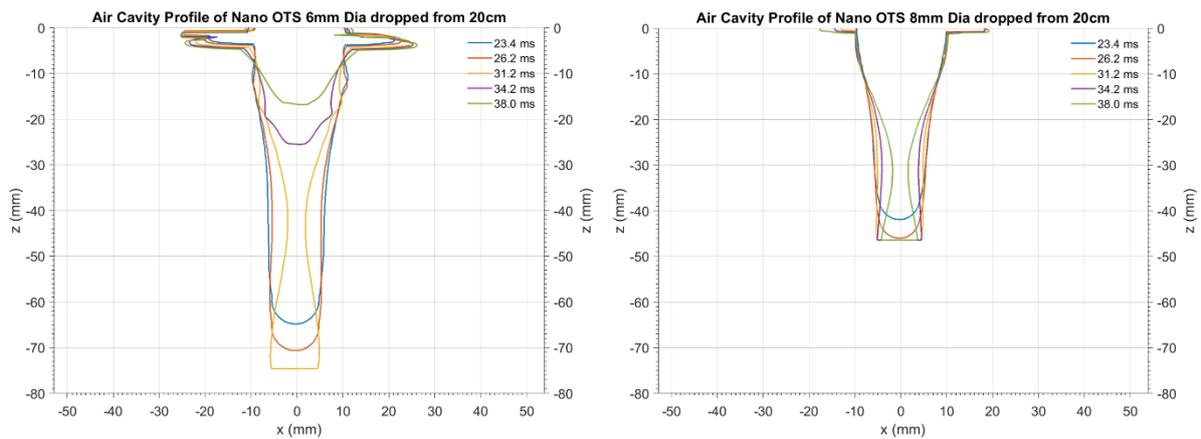

**Figure 11.** Temporal evolution of cavity profiles for Nano OTS dropped from 20 cm for the various sphere diameters.

### 3.4.3. Effect of LIS

The interfacial tension between silicone oil and water is significantly lower than that between water and air. In other words, LIS exhibit lower interfacial energy compared to SHS, indicating reduced resistance to spreading. This enhanced spreading reduces the interfacial energy barrier, minimizes splashing, and facilitates greater energy dissipation through the lubricant layer.

The air cavity dynamics for the micro-nano, and nano LIS surfaces display significant variations in cavity formation and collapse compared to other cases, driven by the presence of the lubricant layer that modifies interfacial energy dissipation and wetting behavior. The image sequence captures the micro-nano LIS impact dynamics with varying diameters (4, 6, and 8 mm) from a height of 20 cm (see figure 12). Each row corresponds to a different diameter, and the columns represent the cavity evolution over time post-impact, ranging from the initial contact to 38.4 ms. Across all sizes (diameter), the sequence illustrates typical deformation behaviors including spreading, cavity formation, and subsequent retraction. Larger diameter exhibits more pronounced deformation, greater cavity depth, and more significant rebound effects compared to smaller droplets. Notably, the 8 mm spherical ball demonstrates stronger oscillations and prolonged interaction with the surface, while the 4 mm sphere shows a relatively quicker recovery. These observations highlight the influence of droplet size on impact dynamics and energy dissipation on micro-nano structured LIS surfaces. The comparison across different scales and surface treatments reveals that the lubricant layer on micro-nano-structured surfaces substantially reduces cavity depth and disrupts stable cavity formation, especially for larger spheres. A MATLAB-generated plot (see figure 13) illustrates the temporal variation of the cavity profile. For heights of 5 and 40 cm, refer to supporting image S5.A.

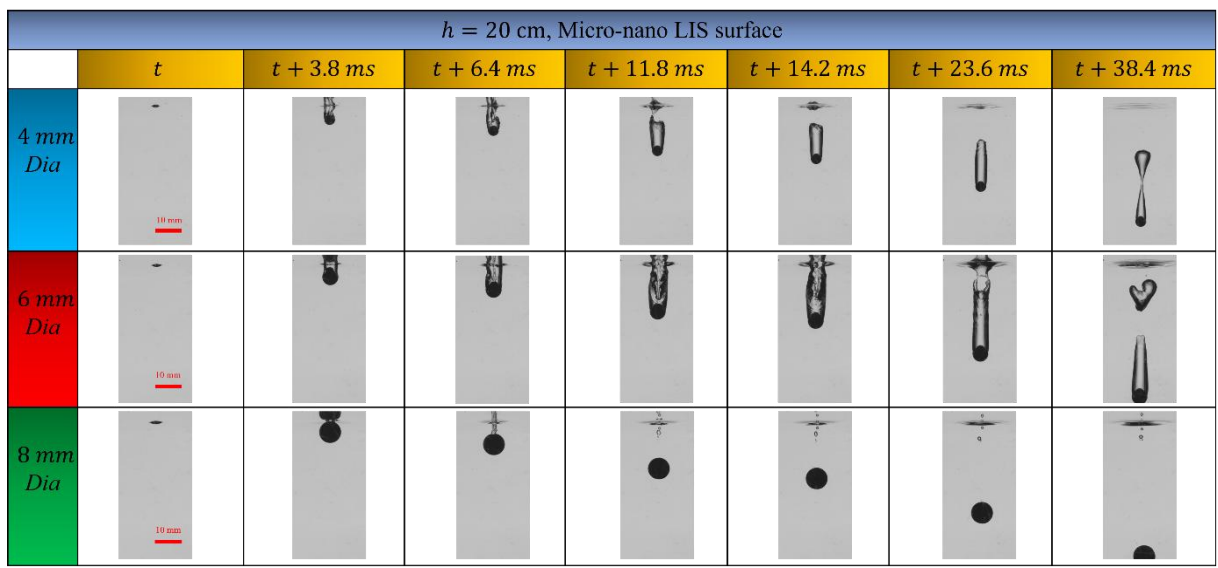

**Figure 12.** Quasi-static, and shallow seal impact cavity. High-speed imaging of micro-nano textured LIS aluminium sphere ball (4,6, and 8 mm dia) impact on liquid from various heights.

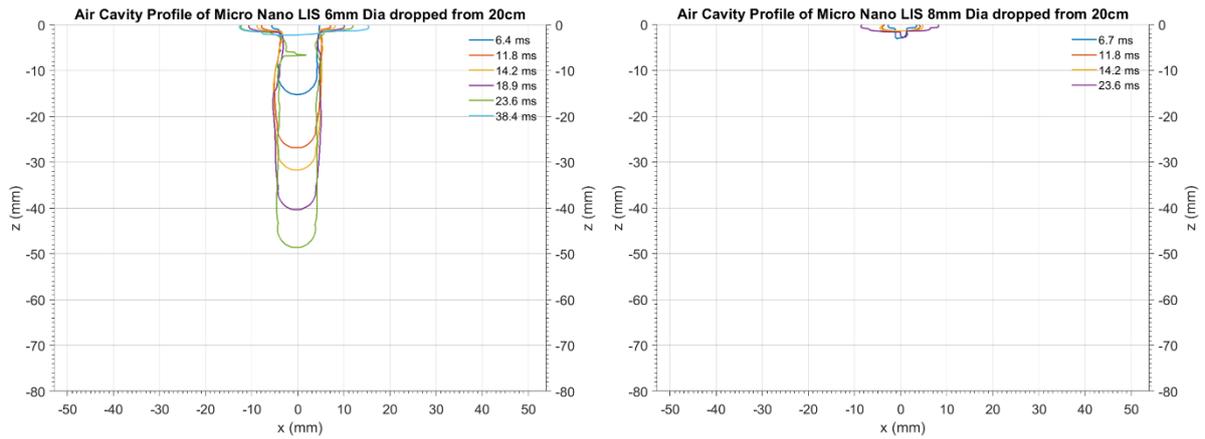

**Figure 13.** Temporal evolution of cavity profiles for micro-nano LIS dropped from 20 cm for the various sphere diameters.

The sequential high-speed imaging of spheres with diameters of 4, 6, and 8 mm impacting a nano LIS from a 20 cm height reveals the size-dependent impact dynamics and cavity formation mechanisms. Additionally, for the height of 5 and 40 cm, refer to supporting image S6.A. Followed by a MATLAB-generated temporal variation of the cavity profile. For the 6 mm sphere, the air cavity formed on the nano LIS surface is notably shallow and rapidly collapses within approximately 10 ms, indicating that the lubricant layer promotes higher energy dissipation and inhibits deep cavity formation (refer to Figure 14). The radial expansion is minimal, and the cavity does not develop into the classical deep-penetrating structure observed in non-lubricated surfaces, suggesting that the viscous and capillary forces dominate the impact dynamics, effectively damping cavity growth. The 8 mm sphere produces the deepest and most stable cavity, with a substantial liquid column ejection and delayed pinch-off, highlighting that the larger mass and inertia can overcome the damping effects of the lubricant layer, leading to more sustained cavity growth and a prolonged collapse sequence. The lubricant layer appears to delay cavity pinch-off, as evidenced by the persistent cavity depth observed up to ~24 ms. The larger mass and higher impact inertia likely overcome some of the dissipative effects of the lubricant layer, enabling deeper penetration and sustained cavity evolution. Across all cases, the LIS surface demonstrates its ability to significantly dampen cavity dynamics compared to non-lubricated surfaces, with the higher spheres experiencing greater suppression of cavity depth. The sequence underscores the critical influence of sphere size on the interplay between inertia, surface energy, and viscous dissipation in governing droplet impact behavior on LIS.

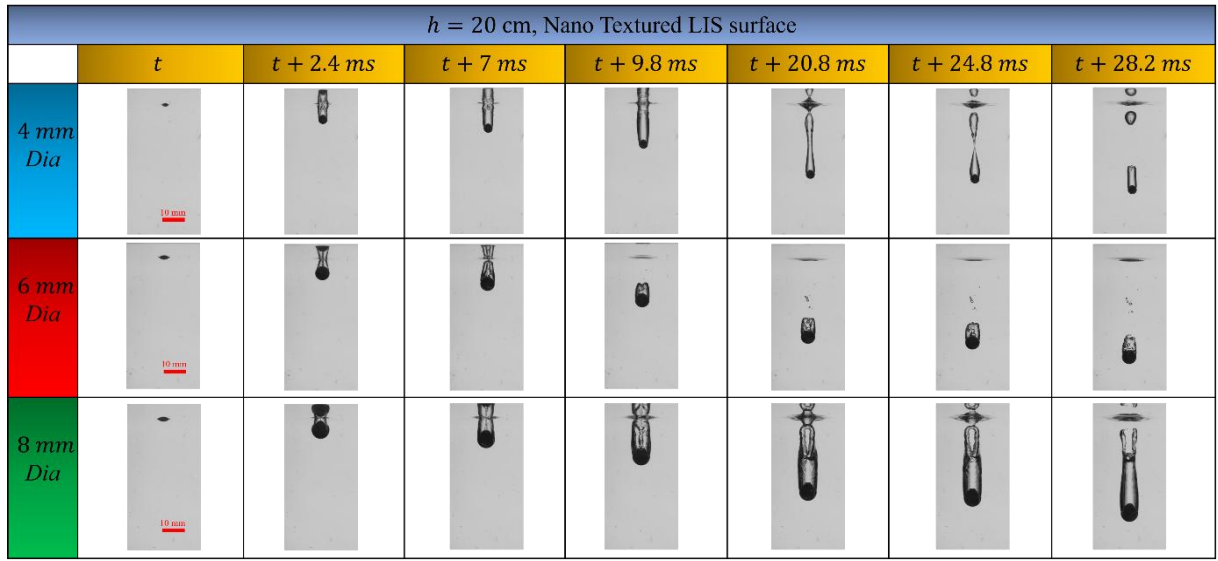

**Figure 14.** Deep seal and inverse LIS impact cavity. High-speed imaging of control and nano-textured functionalized aluminium 8mm sphere ball impact on liquid from 20 cm height.

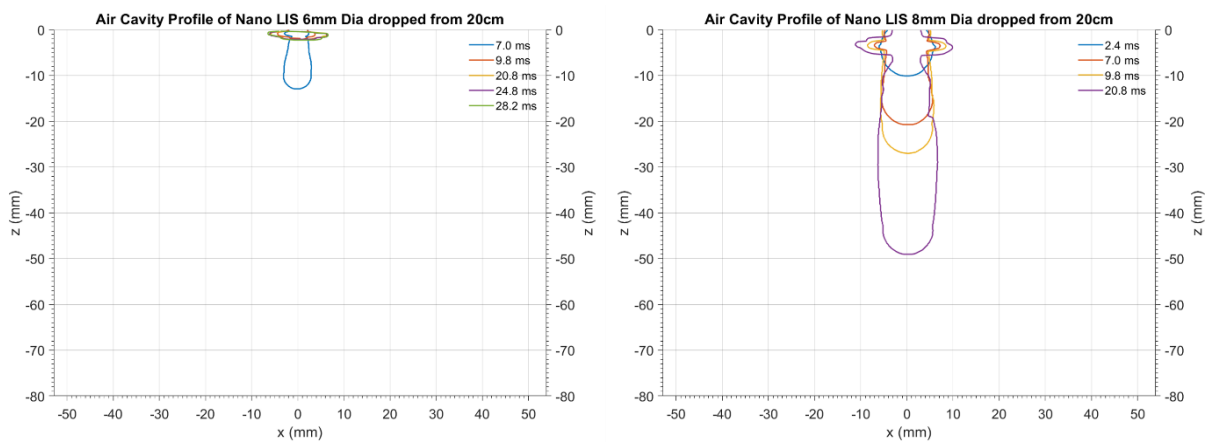

**Figure 15.** Temporal evolution of cavity profiles for Nano LIS dropped from 20 cm for the various sphere diameters.

Qualitatively, it is evident that LIS tend to produce quasi-static and shallow seal impact cavities. Air entrapment is minimal, with only tiny bubbles attached to the sphere after the pinch-off.

**3.5. Trajectory analysis**

The sphere trajectory plots (see figure 16) detail the depth-time evolution for 6 mm and 8 mm diameter spheres impacting variously treated surfaces from a 20 cm drop height. For both sphere sizes, the trajectories reveal that spheres on lubricant-infused surfaces (Nano LIS and Micro-nano LIS) penetrate deeper and maintain higher velocities over longer times, reflecting reduced hydrodynamic resistance. The 6 mm sphere trajectories show that the Control and Control + OTS surfaces quickly decelerate, leading to shallower penetration depths, which is indicative of significant energy dissipation through fluid drag and cavity formation resistance. Notably, the Control + OTS case, despite its hydrophobic coating, exhibits a relatively slower descent compared to textured and lubricated surfaces, emphasizing that surface chemistry alone does not sufficiently reduce drag in the absence of surface texturing or lubricant layers.

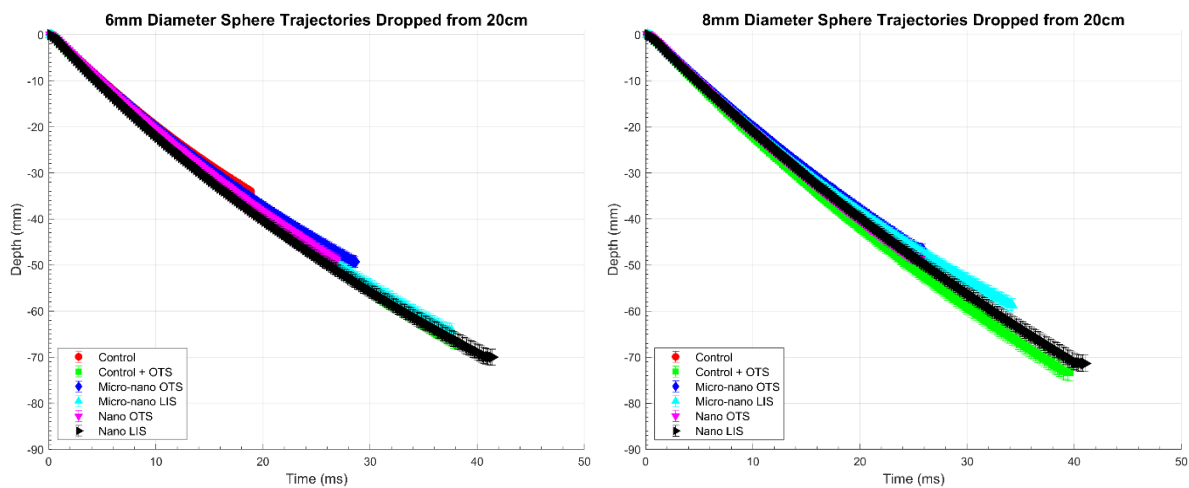

**Figure 16.** Trajectory of a 6, 8 mm diameter sphere dropped from a height of 20 cm.

In the case of the 8 mm sphere, the trends persist but the larger mass results in deeper penetration across all surfaces due to greater momentum. However, the micro-nano OTS surface exhibits considerable deceleration, suggesting that although texturing reduces drag to an extent, it is not as effective as LIS in promoting sustained descent. The LIS surfaces, especially the Nano LIS, consistently demonstrate the deepest and fastest trajectories, highlighting their superior drag-reduction capabilities attributed to the lubricant layer which promotes slip and minimizes direct solid-liquid contact. These observations confirm that the combination of hierarchical roughness and lubricant infusion synergistically enhances drag reduction, allowing spheres to travel deeper and more efficiently through the fluid medium.

### 3.6. Velocity and kinetic energy analysis

The velocity with time plots for both 6 mm and 8 mm diameter spheres dropped from a height of 20 cm provide detailed insights into the deceleration behavior influenced by surface properties (see figure 17). To enable a clear distinction between the samples, a zoomed-in plot was presented, since the data points exhibit considerable overlap during the initial time interval. For the 6 mm spheres (see figure 17, top row), the initial velocity is approximately 2 m/s, and a clear divergence in deceleration rates is observed across various surfaces. The control sphere experiences the most rapid deceleration due to significant fluid resistance. The control with functionalisation surface slightly mitigates the drag force due to the coating, maintaining a higher velocity for a longer duration compared to the untreated surface. Upon introducing the texture, it was observed that the nano-OTS surface exhibited a slightly higher velocity compared to the micro-nano-OTS surface. However, the difference in magnitude was minimal, likely due to the intrinsic characteristics of the surface texture. Notably, spheres interacting with LIS, especially the Nano LIS, sustain the highest velocities throughout the measured time window, demonstrating the superior drag-reduction effect of the lubricant layer and nanoscale texturing.

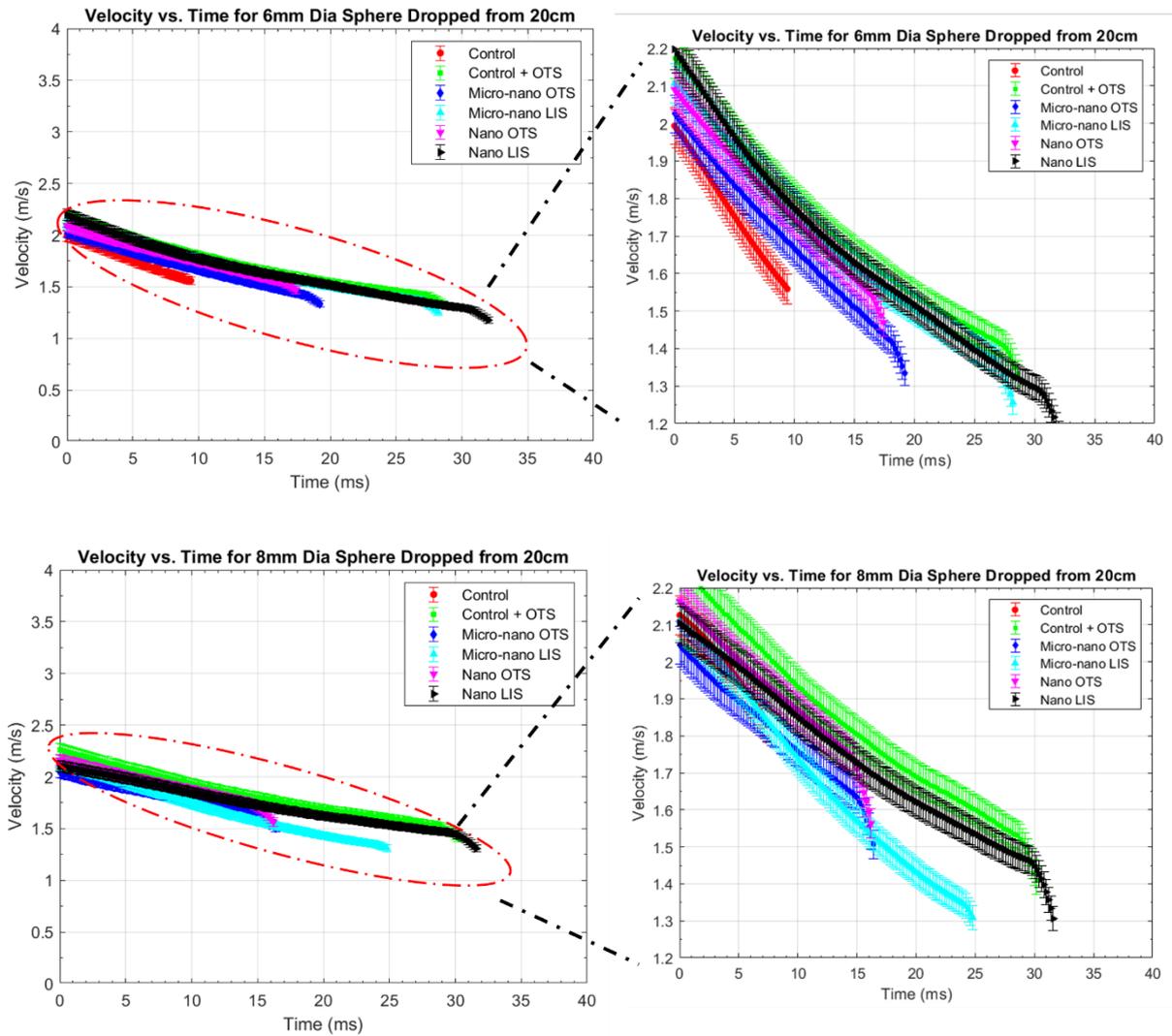

**Figure 17.** Velocity vs time profiles for spheres dropped from a height of 20 cm, arranged row-wise for 6- and 8-mm diameters, respectively.

For the 8 mm spheres (see figure 17, bottom row), the trends remain consistent, but the larger mass results in relatively higher initial velocities, approximately 2.2 m/s, and an overall slower rate of deceleration across all surfaces due to increased inertia. Particularly within the first few milliseconds, resulting in overlapping data points that make differentiation challenging in this region. To address this, a zoomed-in view was provided in the study to clearly distinguish the minute changes between the samples at the early time scale. As time progresses, the velocity decay patterns begin to diverge more prominently, indicating varying resistive effects and energy dissipation characteristics associated with the different surface treatments. SHS, the micro-nano OTS exhibited lower velocities at the initial time scale, whereas the nano OTS surfaces initially demonstrated higher velocities, which subsequently dropped over time. The Micro-nano LIS shows the sharpest drop in velocity, indicating that despite the hierarchical texturing, the lubricant layer plays a critical role in minimizing drag. The Nano LIS again exhibits the least velocity loss, underscoring its efficiency in maintaining slip and reducing fluid-surface interaction. These velocity-time profiles quantitatively highlight the importance of lubricant infusion and nanoscale texturing in enhancing hydrodynamic performance and delaying energy dissipation during sphere descent.

The provided graphs (see figure 18) illustrate the normalized velocity with time for 6- and 8-mm diameter spherical balls dropped from a height of 20 cm on various treated surfaces. To conduct a non-dimensional analysis of the plot, the maximum velocity and corresponding time were systematically calculated. In both cases, the velocity decreases progressively due to fluid drag and cavity formation resistance, but the rate of deceleration is distinctly influenced by surface properties, i.e., wettability. For the 6 mm sphere, the control surface exhibits the steepest velocity decay, indicating the highest drag, likely due to greater fluid adhesion and cavity resistance. This trend is consistent in the 8 mm sphere case, although the velocity decay rates are comparatively less steep across all surfaces due to the increased mass and momentum of the larger sphere. These results qualitatively confirm that the combination of hierarchical roughness and lubricant impregnation significantly minimizes drag forces during impact and subsequent cavity penetration.

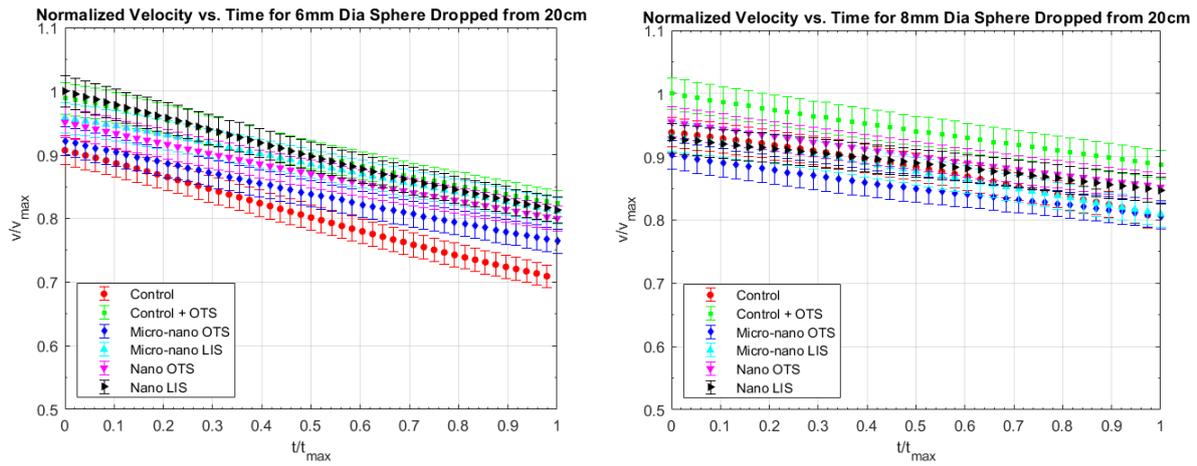

**Figure 18.** Normalized velocity vs normalized time for a 6 and 8mm diameter sphere dropped from a height of 20 cm.

### 3.6.1. Kinetic energy

The plot (refer figure 19) below shows the kinetic energy evolution over time for 6- and 8-mm spheres dropped from 20 cm onto surfaces with varying wettability and microstructure. The kinetic energy dissipation is tracked over time for different surface conditions. For the 6 mm sphere, the kinetic energy dissipates more rapidly on all surfaces except the control one, indicating higher energy loss due to enhanced wetting resistance and possibly increased interfacial interactions on structured surfaces. The nano LIS shows the most significant energy dissipation rate. For the 8 mm sphere, the overall kinetic energy values are higher due to the increased mass, but a similar trend of energy reduction hierarchy is observed. Notably, the control with OTS functionalisation surface retains kinetic energy longer, indicating lower energy dissipation compared to LIS-treated surfaces. The broader spread of the data in the 8 mm case suggests greater variability in impact dynamics, possibly due to increased gravitational potential and higher impact forces. The differences across surfaces reflect the complex interplay between surface texture, liquid-infused layers, and hydrophobic coatings in modulating fluid structure interactions and energy dissipation pathways.

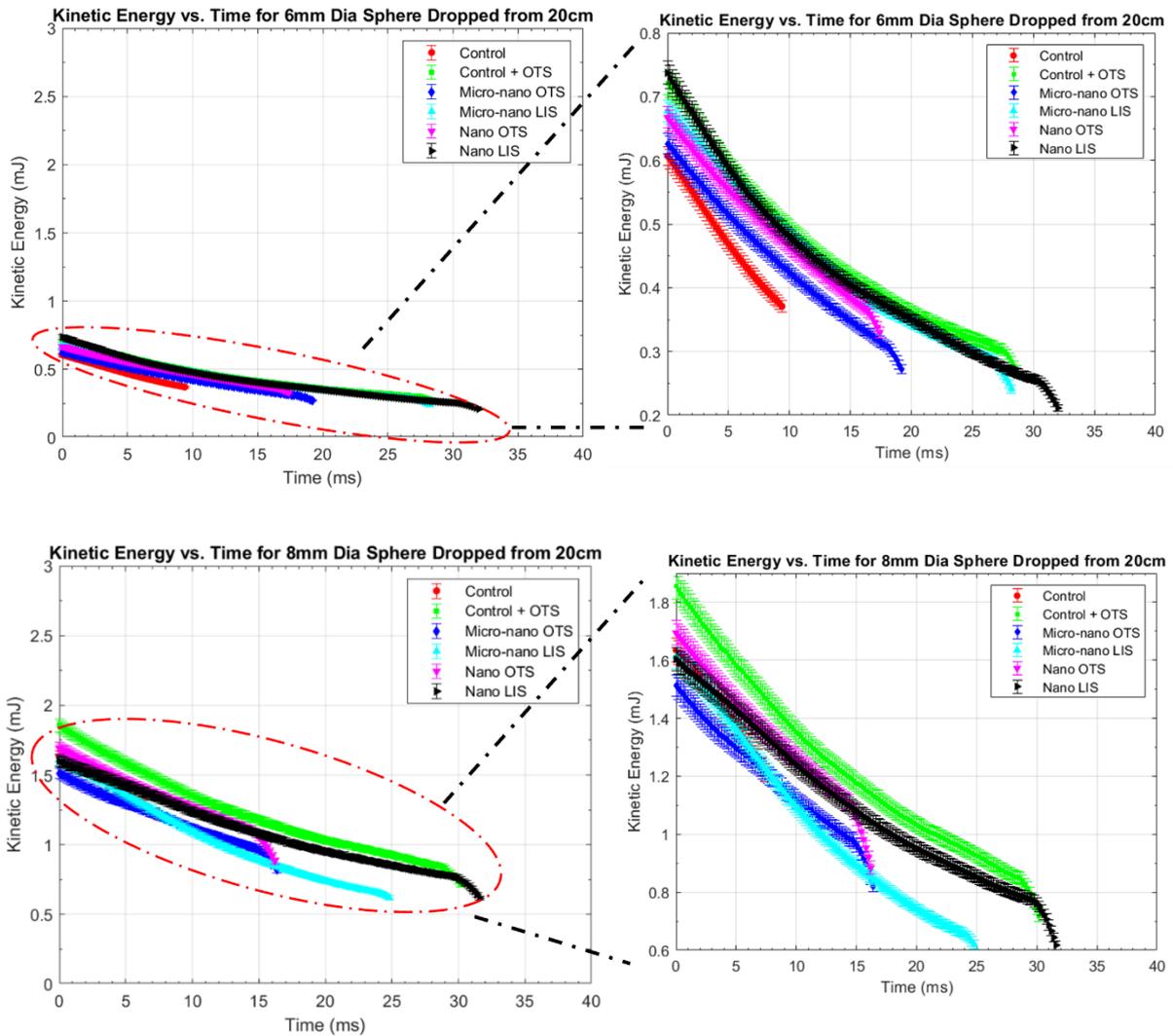

**Figure 19.** Kinetic energy vs time profiles for spheres dropped from a height of 20 cm, arranged row-wise for 6- and 8-mm diameters, respectively

For the 6 mm sphere (see figure 19, top row), the kinetic energy decreases from 0.7 mJ to 0.4 mJ within 10 ms after water entry. Similarly, for the 8 mm sphere (refer figure 19, bottom row), the kinetic energy declines from 1.7 mJ to 1.2 mJ within 10 ms across nearly all tested surfaces. The deceleration phase exhibits significant kinetic energy dissipation, with greater losses observed for the 8 mm sphere upon water impact.

The graphs (see image 20) illustrate the normalized kinetic energy with time for 6- and 8-mm diameter spheres dropped from a height of 20 cm onto surfaces with varying roughness, chemical functionalization, and lubrication. The kinetic energy dissipation is a direct indicator of momentum loss, primarily due to hydrodynamic drag and energy transfer to the surrounding liquid. For both sphere sizes, the control surface with functionalisation consistently exhibits the fastest energy decay. In contrast, surfaces with lubricant impregnation, particularly the Nano LIS and Micro-nano LIS, maintain higher normalized kinetic energy over time. Textured surfaces without lubricant (Micro-nano OTS and Nano OTS) show intermediate dissipation rates, where nano and micro-nano roughness contribute to partial drag reduction but are insufficient to match the energy retention offered by LIS. Overall, the data quantitatively demonstrate that LIS significantly suppresses kinetic energy loss, thereby offering substantial drag reduction during water entry dynamics. For the 6 mm spheres, the kinetic energy loss ranges from 30% to 50%, whereas for the 8 mm spheres, the loss is between 20% and 30%. These results demonstrate that both sphere size and surface roughness significantly influence the hydrodynamic behavior and cavity evolution. The dissipated energy is transferred to the surrounding fluid, driving radial and axial fluid motion against resistive forces, including surface tension, buoyancy effects, and hydrodynamic resistance.

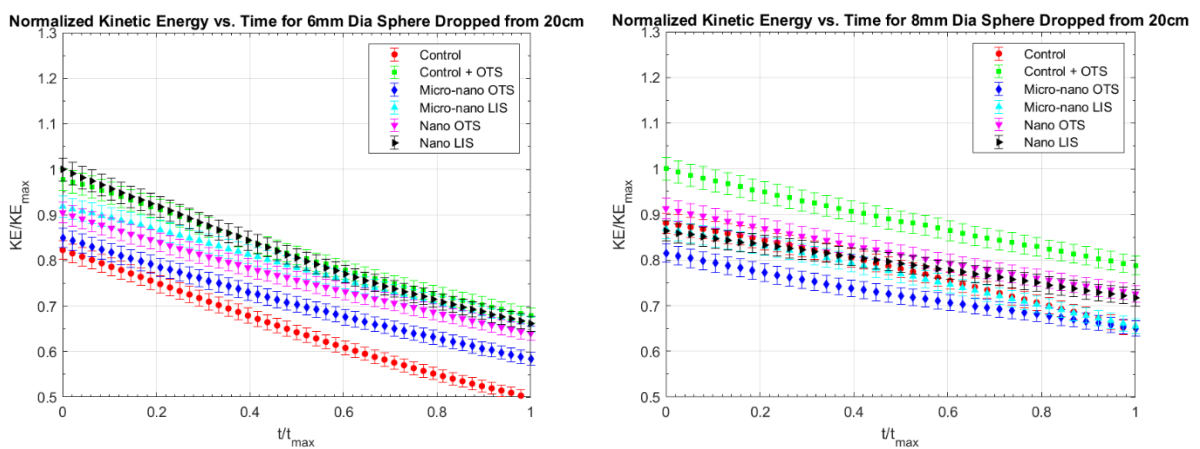

**Figure 20.** Normalized kinetic energy vs normalized time for a 6 and 8mm diameter sphere dropped from a height of 20 cm.

### 3.7. Pinch off time, depth, and sphere location

The pinch-off time, depth, and sphere location are critical parameters in studying impact behavior. These factors influence the cavity formation, detachment, and subsequent motion upon interaction with the surface. Understanding their relationship provides insight into fluid-structure interactions. This analysis examines how these variables vary under different experimental conditions. The plot (see figure 21) shows the variation in pinch-off time for spheres of different sizes (4, 6, and 8 mm) and release heights (5, 20, and 40 cm) impacting various surface types. The pinch-off time, which reflects the duration before the cavity detaches from the surface, is significantly influenced by both the surface texture and chemical modification. Notably, Micro-nano OTS and Nano OTS surfaces exhibit substantially higher pinch-off times across most conditions, indicating increased interaction and retention of the cavity, likely due to the texture. In contrast, the control and control with OTS surfaces show the shortest pinch-off durations. The LIS surface lies between the control and textured surface. The trend also highlights that larger cavities and higher release heights generally result in longer pinch-off times, especially on textured and lubricated surfaces. The observations show that there is an increase in pinch-off time with an increase in the impact velocity (see Figure 21). A non-linear trend observed for the variation in the H, as well as pinch off time, is found to be associated with the impact velocity.

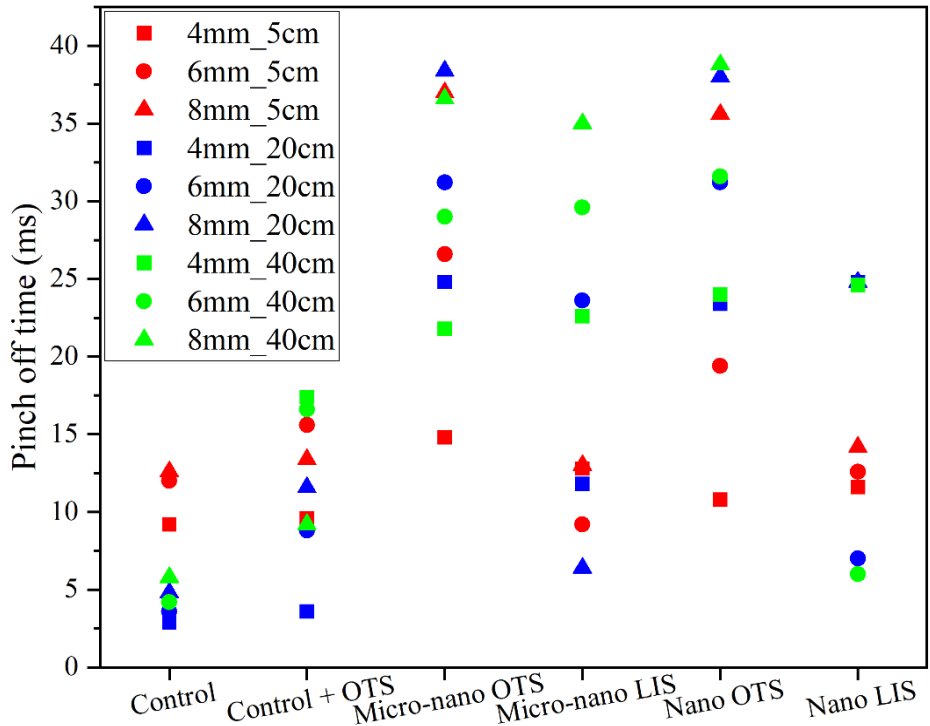

**Figure 21.** Pinch off time of all materials of different diameters dropped from various heights

The plot (see figure 22) illustrates the relationship between the impact velocity of a 6 mm diameter sphere ball and two key parameters: the pinch-off depth (in black, left axis) and the sphere location at pinch-off time (in red, right axis). Four different surface conditions are compared: MN OTS, MN LIS, Nano OTS, and Nano LIS. As the velocity increases, the pinch-off depth increases or remains almost the same, except for nano LIS. Indicating a deep seal cavity formation on textured surfaces. In the shallow seal case, the pinch-off depth (h) is dynamic in nature with an increase in the impact velocity. However, for the deep-seal cases, it is found that the pinch-off depth increases with an increase in the impact velocity. Conversely, the sphere location at pinch-off time tends to increase with velocity, showing deeper sphere positions at the moment of pinch-off. Notably, the Nano LIS surfaces exhibit the deepest pinch-off depths at lower velocities and the shallowest at higher velocities, suggesting a distinct surface-liquid interaction compared to the other surfaces. The trends highlight the significant influence of both surface microstructure and lubrication in underwater dynamics.

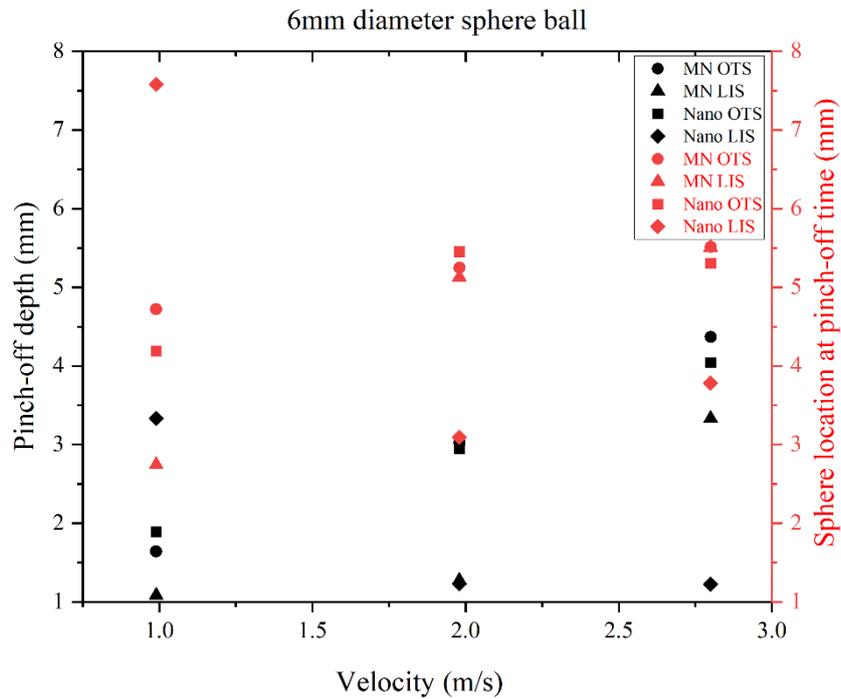

**Figure 22.** Pinch off depth (h) and sphere location at pinch off time (H) for 6mm spherical balls of SHS and LIS dropped from various heights

### 3.8. Various dimensional numbers

The presented figures (refer figure 23) illustrate the impact dynamics of a 6 mm diameter spherical ball on various surface types, characterized by different wetting properties and textures, as a function of Weber number, Froude number, and Bond number. The datasets include control, Control OTS, Micro-Nano OTS, Micro-Nano LIS, Nano OTS, and Nano LIS surfaces. It has been shown from the analysis carried out by Aristoff et al.[22] based on the analytical model that the cavity pinch-off depth (h), sphere location at the pinch-off time (H), and pinch-off time have linear variation with deep seal cavities. In the Weber number plots, both the normalized (h/d) and (H/d) generally increase with increasing Weber number, with distinct variations across surface types except for nano LIS. The value of the Weber number is greater than one, with the dominance of the inertial forces over the surface tension forces.

Notably, surfaces like Micro-Nano LIS exhibit higher penetration compared to the control, indicating reduced resistance and enhanced energy retention during impact. Similarly, in the Froude number-based analysis, a consistent trend of increased h/d and H/d with Froude number is observed, with the LIS surfaces showing superior dynamic responses. The Bond number plots, however, demonstrate minimal variation, indicating that gravitational effects remain relatively constant across the experiments due to the fixed sphere size. Overall, the results emphasize the significant influence of surface texturing and lubrication on the impact dynamics of submerged spheres.

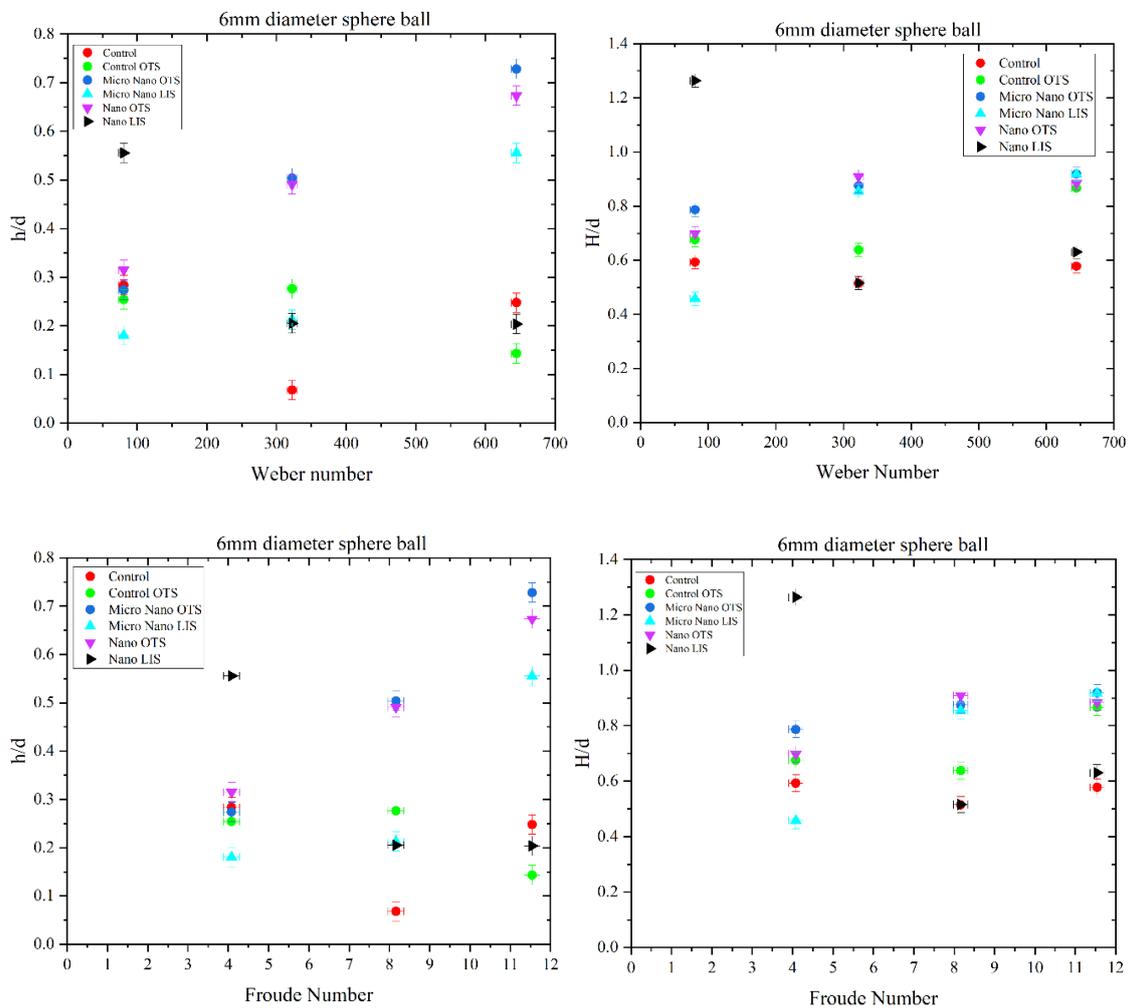

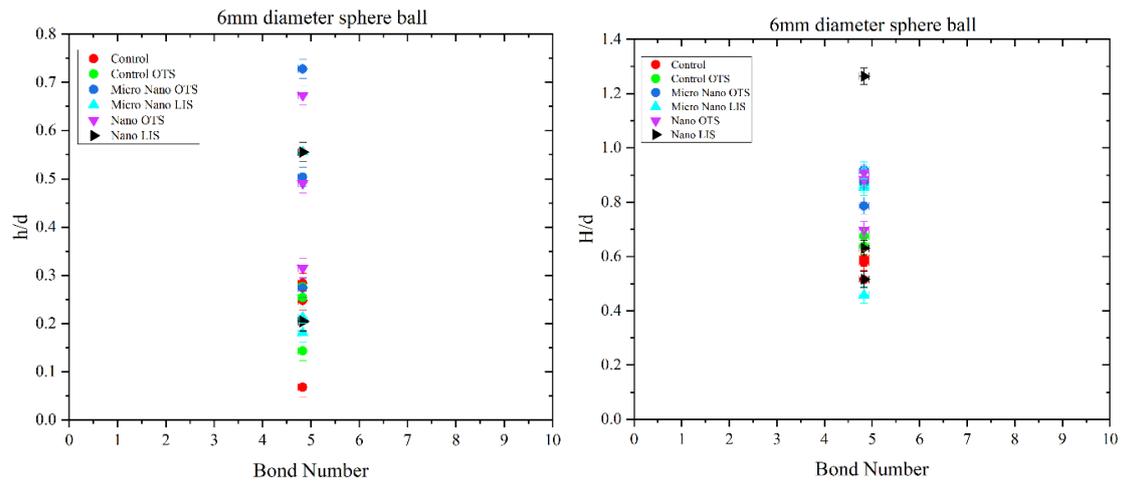

**Figure 23.** Variations of the dimensionless pinch-off depth (h) with Weber number (We), Froude number (Fr), and Bond number (Bo), respectively. And variation of sphere location at pinch-off time (H) with Weber number, Froude number, and Bond number, respectively.

## 4. Conclusions

This study was conducted to investigate lubricant drainage in a dynamic environment. The results indicate that LIS shows the sharpest drop in velocity, indicating that despite the hierarchical texturing, the lubricant layer plays a critical role in minimizing drag. Further analysis was carried out for different regimes, such as a quasi-static, shallow seal, and deep seal impact cavity. Additional investigations were performed, such as velocity, kinetic energy, pinch off depth, pinch off time, sphere location at pinch off time, various non-dimensional parameters with the impact velocity, and various surface types. The pinch-off phenomenon and Worthington jet height have been observed on all surfaces; however, they are highest on textured ones. The study demonstrates that hierarchical surface structuring combined with lubricant infusion dramatically alters the impact dynamics of spherical bodies. LIS, particularly those with micro-nano textures, effectively dissipate kinetic energy and suppress cavity formation, outperforming both smooth and OTS-functionalized rough surfaces. The analysis of the trajectory data confirmed no significant deviations in the plotted trends, indicating that the lubricant remains stable within the LIS with no signs of degradation or adverse interactions. Furthermore, the velocity profile exhibited smooth and consistent behavior, with no abrupt fluctuations, suggesting that there is no measurable lubricant drainage from the surface under the observed conditions. These findings provide qualitative validation of the lubricant's stability and retention, supporting its continued effectiveness in the system. These findings highlight the potential of LIS in applications requiring impact damping, splash suppression, and energy dissipation. The ultimate goal of this study is to develop highly energy-efficient products. Future efforts will focus on optimizing the relevant parameters to enhance performance further. Future research on solid impact behavior would be investigating the effects of object geometry, size, impact angle, and velocity on interfacial interactions with engineered surfaces. Scaling analysis and non-dimensional representations of these phenomena could further generalize the findings, facilitating predictive modeling for broader industrial applications. Water entry dynamics studies comparing SHS and LIS surfaces demonstrated superior performance of textured surfaces across multiple metrics: reduced drag, enhanced energy dissipation, and cavity dynamics. Systematic analysis of trajectory, cavity dynamics, pinch-off phenomena, Worthington jet formation, velocity and normalized velocity, kinetic energy, pinch-off depth, and sphere location at pinch-off time. Time-resolved kinetic energy profiles revealed that micro-nano textured LIS optimally suppresses cavity formation while maximizing energy dissipation. These findings establish LIS, particularly hierarchical designs, as superior to both smooth and SHS for impact-related applications requiring splash suppression and energy damping.

Future research on solid impact behavior should explore the influence of object geometry, size, impact angle, and velocity on interfacial interactions with various engineered surfaces. Scaling analysis and non-dimensional representations of these phenomena could further generalize the findings, enabling predictive modeling across a wider range of industrial conditions. A detailed examination of different impact regimes, such as quasi-static, shallow seal, and deep seal cavity formation, would elucidate the underlying fluid-structure interactions and energy dissipation mechanisms. Additionally, the development of advanced experimental setups capable of high-resolution visualization of lubricant drainage under dynamic conditions remains a critical challenge. Such investigations would not only validate theoretical predictions but also provide empirical benchmarks for designing next-generation liquid-repellent surfaces.

## 5. Materials and Methods

### 5.1 Experimental procedure

The experimental procedures (see figure 24) include the fabrication process of a LIS using an aluminum (Al) sphere. The process begins with cleaning with solvents, followed by a CVD treatment to functionalize the surface. The treated surface then undergoes wettability measurement to assess surface modification efficacy. Post-CVD, the surface undergoes acid-based nano-texturing (1M) or micro-nano texturing (2.5M) to develop hierarchical roughness, followed by boiling water treatment. These textured surfaces are then immersed in the lubricant oil bath using a precision dip coating, resulting in the formation of the LIS with desirable wetting and functional properties. Once the sample was prepared, the experiments were conducted using a solid-impact setup consisting of multiple components, as shown in Figure 6.1. The experiments were performed inside a transparent plexiglass chamber. A custom-fabricated plexiglass tank was used, equipped with a high-speed camera and an appropriate light source to enable video recording. The sample was securely held by a solenoid valve and released from varying heights to study the impact dynamics.

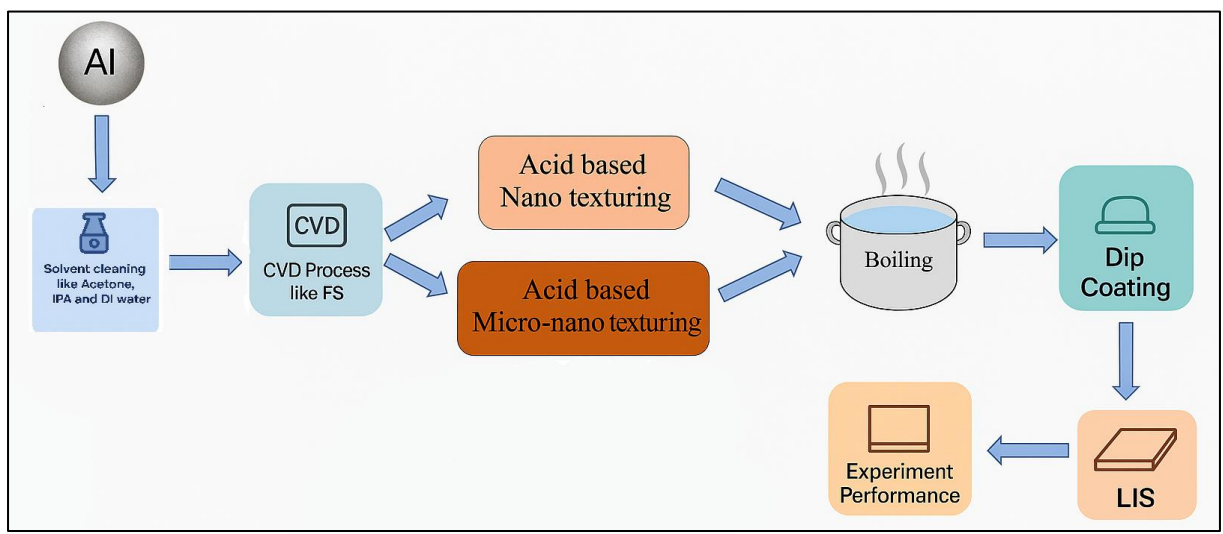

**Figure 24.** Experimental Workflow Schematic for Water Entry Dynamics Experiments

The process begins with an aluminum (Al) sphere, which undergoes a cleaning step to prepare the surface. The sphere is further processed through a Chemical Vapor Deposition (CVD) process, followed by cleaning. After the CVD treatment, two texturing methods can be employed: acid-based nano-texturing and acid-based micro-nano texturing. These texturing methods create specific surface patterns on the sphere. The textured surface is then processed further, i.e., boiling to generate nano-texturing. This structure is subsequently dip-coated to make LIS surface.

## 5.2 Surface chemical functionalization

Solid surfaces with and without texturing were chemically altered using a widely known silane process. For this objective, a low surface energy silane OTS (Trichloro (octadecyl) silane, Sigma Aldrich) was used. Aluminum sphere balls, and strips were first plasma cleaned for 10 minutes in an RF plasma cleaner (Harrick Plasma, Model PDC-002-HP) in an oxygen-rich environment at 200 mTorr pressure. Subsequently, samples were submerged in toluene (ACS grade 99%, Sigma Aldrich) containing 0.5 wt% OTS. An emulsion of 1 wt% DI water (Millipore, 18.2MΩ) in toluene was made separately and added to the OTS-Toluene solution in equal volumes. The water in the toluene emulsion was synthesized by sonicating with a high-energy probe (Sonics 750W sonicator) for 90 seconds at 70% power, followed by 10 minutes in a bath sonicator (Branson). Avoid employing probe and bath sonication on silicon/aluminum samples with nanostructures, as this can destroy them. Samples were immersed in a mixture of OTS, Toluene, and DI water for 20 minutes to complete the reaction. Once the chemical reaction was completed, the samples were bath sonicated for three minutes separately in Acetone (ACS grade, 99%-Sigma Aldrich) and Isopropyl Alcohol (ACS grade, 99%-Sigma Aldrich), followed by a thorough rinsing with DI water. These cleaning methods have been employed to remove physically absorbed silanes from sample surfaces.

**5.3 Acid Etching Process:**

The aluminum sphere balls were procured locally, while hydrochloric acid (HCl, 37% concentration) was purchased from Merck India Pvt. Ltd. To ensure surface cleanliness, the substrates were ultrasonically cleaned in sequential baths of acetone, isopropanol, and deionized (DI) water. Hydrochloric acid-based chemical etching (2.5M HCl solutions) was used to generate various microstructures on these aluminum samples. After 15 minutes of etching, the samples were removed to evaluate the influence of reaction time on microfeature development. Following acid treatment, the samples were rinsed thoroughly with DI water. Subsequently, the cleaned surfaces were immersed in boiling water for 20 minutes to induce the formation of nanopetal-like structures. Finally, the samples were dried in a convection oven and stored in a vacuum desiccator to prevent contamination before further characterization or processing. For nano-texturing the aluminum surface, a similar two-step process was employed. First, the surface was etched using a 1M HCl solution to remove impurities and oxides, ensuring a clean and reactive base. This was followed by a boiling water treatment, which promoted the growth of a nanostructure, creating a non-uniform textured surface. This method enhances surface roughness and improves functional properties.

**5.4 Microscopic Visualization:**

The textured surfaces were characterized using a field emission scanning electron microscope (FESEM) (Sigma 300-Carl-Zeiss). A typical image was captured at a voltage of around 5kV and a magnification of approximately 1500x.

**5.5 Contact angle measurement:**

A contact angle goniometer (Rame-Hart model 500-U1) was employed to measure the equilibrium, receding, and advancing contact angles on various surfaces in air, water, and cyclopentane environments. Test samples were immersed in a clear quartz cell filled with liquid to measure contact angles in a liquid environment. The size of probe liquids was fixed below their respective capillary lengths (~4 μL). A dispensing and retraction rate of 0.2 μL/s was employed to evaluate the advancing and receding angles. To compute the average and standard deviation of observed quantities, at least ten measurements have been recorded from various points on the samples.

**5.6 Dip coating process:**

The lubricant was impregnated into the textured OTS functionalized surfaces using the dip coating methodology to produce LIS surfaces. Regardless of surface roughness properties, the dip coating process prevents the formation of a thick lubricating layer. This method was carried out using a precision dip coater (Biolin Scientific: KSV NIMA multi vessel). A traditional vessel was filled with lubricating oil after being carefully cleaned using solvents like acetone and isopropanol. A functionalized rough surface mount on the cantilever allows it to be immersed in the lubricant at a specified Y speed (mm/min). Provide the appropriate withdrawal velocity after the sample is completely submerged, then carry out the experiment.

AUTHOR INFORMATION

**Corresponding Author**

* Email: arindam@iitgoa.ac.in

**ORCID ID**

Abhishek Mund: 0000-0003-2924-1804

Shubham S Ganar: 0000-0003-3767-2460

Arindam Das: 0000-0002-8163-0666

AUTHOR CONTRIBUTION

Abhishek Mund: Designed and performed experiments, Theoretical calculations, Data analysis, Writing original manuscript. Shubham S. Ganar: Experiments performed, Review and editing the manuscript. Arindam Das: Developed the experimental design and concept, Review and editing the manuscript, Supervision. The final document of the article has received the unanimous approval of all writers.

**Notes**

The authors declare no competing financial interest.

ACKNOWLEDGMENT

This research was partially funded by the Anusandhan National Research Framework (ANRF), formerly known as the Science and Engineering Research Board (SERB) Core Research Grant (CRG) with sanction order nos. CRG/2023/008620. The authors thank CoE-PCI (Centre of Excellence-Particulates, Colloids, and Interface) and the School of Mechanical Sciences of IIT Goa for providing the instrument facility and space to perform the experiment. Abhishek Mund and Shubham S. Ganar acknowledge the financial assistance received under the Ministry of Education (MoE), Government of India. We would also like to thank Rahul Khimjibhai Tilwani, Yash Khobragade, and Amey Sunil Bhagwat. We thank anonymous reviewers for their valuable feedback and comments.

**REFERENECS**

**Image S1.A. High-speed imaging (under water) of various diameter control surfaces at different heights.**

The image sequence below illustrates the impact dynamics of spherical balls with varying diameters of 4, 6, and 8 mm released from various heights, such as 5, 20, and 40 cm, onto a liquid surface. For all sizes, the initial contact (at time $t$) triggers surface deformation and cavity formation. At the lowest height, all spheres show minimal splashing and gradually deform into spherical shapes with subtle surface waves. As the height increases to 20 cm, the impact velocity increases, leading to more pronounced deformation and the onset of air entrainment and bubble formation, especially for larger dia. At 40 cm, larger jet height, and shallow cavity formation, particularly in the 6 mm and 8 mm diameter cases, indicating a strong dependence of impact dynamics on both sphere size and drop height. This progression emphasizes the interplay between kinetic energy, inertia, and surface tension in governing the outcome of solid impacts. The clear differences between the three rows highlight the scale dependence of splash dynamics and underscore the importance of diameter in determining impact outcomes on fluid surfaces.

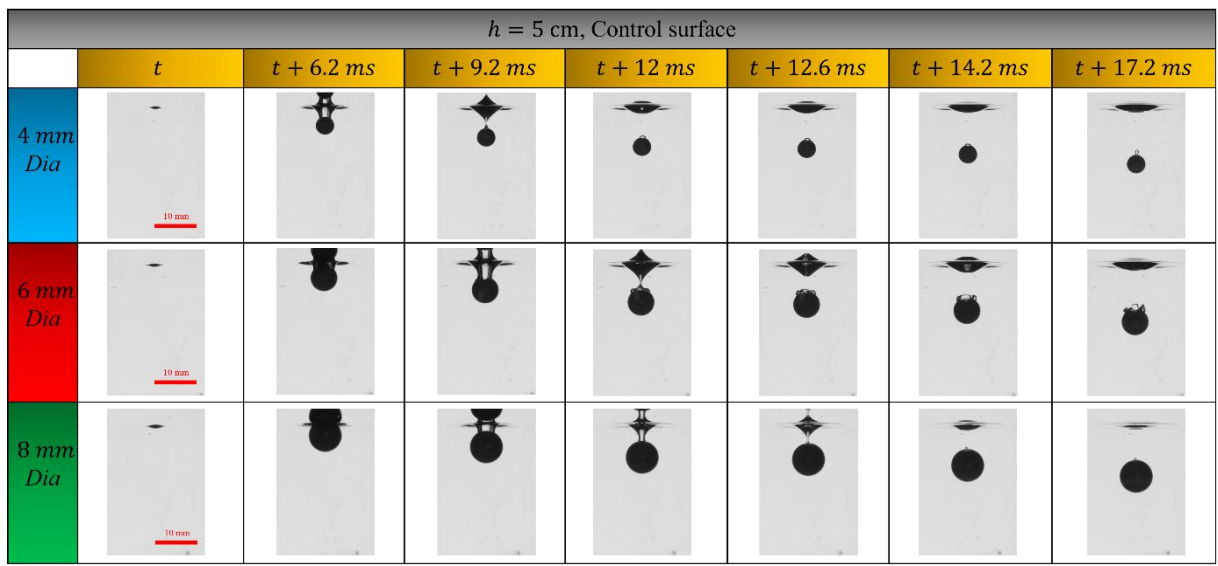

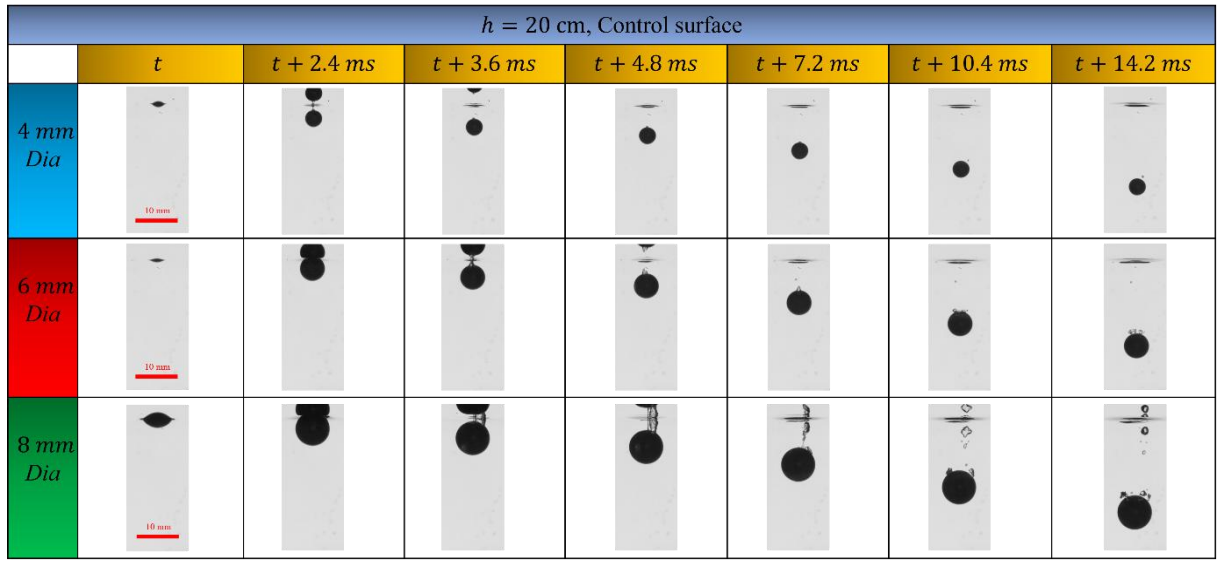

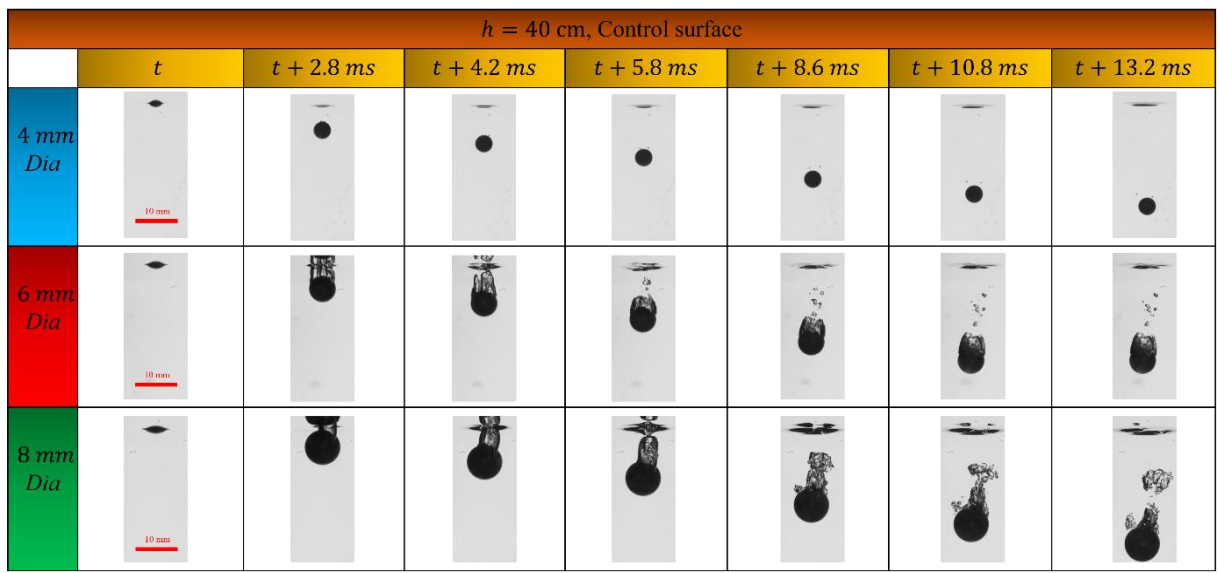

**Image S1.B. High-speed imaging (above the water surface) of various diameter control surfaces at different heights.**

The side-view images depict the post-impact dynamics of spherical balls of varying diameters (4, 6, and 8 mm) impacting the water surface from three different heights (5, 20, and 40 cm). At 5 cm, the ball spread symmetrically with minimal rebound, especially in the smaller sizes. The 8 mm ball shows a low jet. At 20 cm, more vigorous dynamics emerge, with distinct splash patterns in the larger diameter and the formation of vertical jets. The 8 mm ball from this height produces a tall, symmetric Worthington jet that culminates in the detachment of a secondary droplet. At 40 cm, the energy is sufficient to generate tall jets and, with more pronounced jet and cavity dynamics observed in the larger diameters. These observations emphasize the influence of impact energy and ball size on splash morphology, jet height, and satellite droplet formation during and after impact.

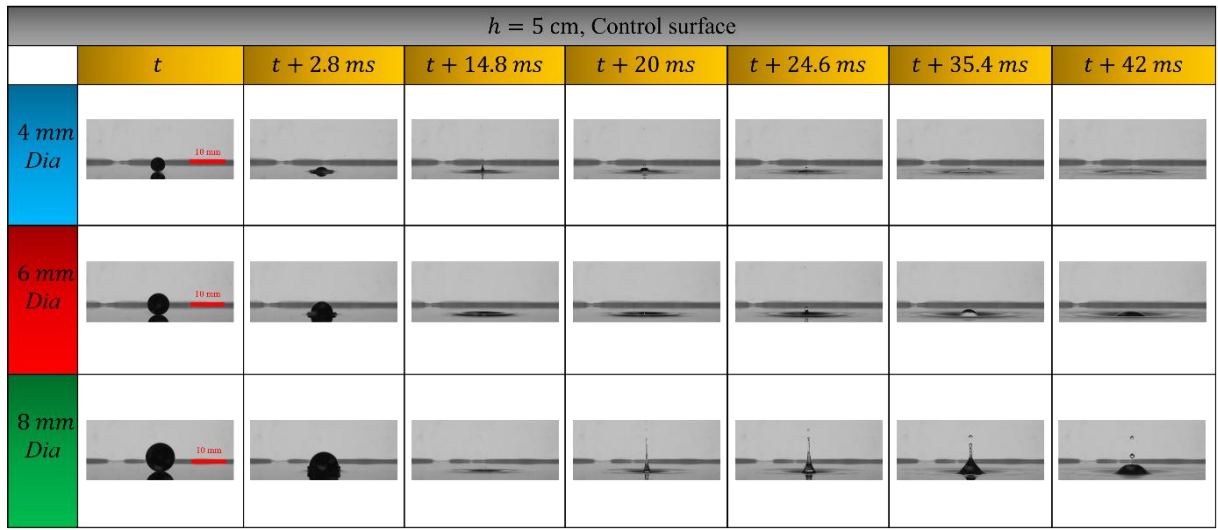

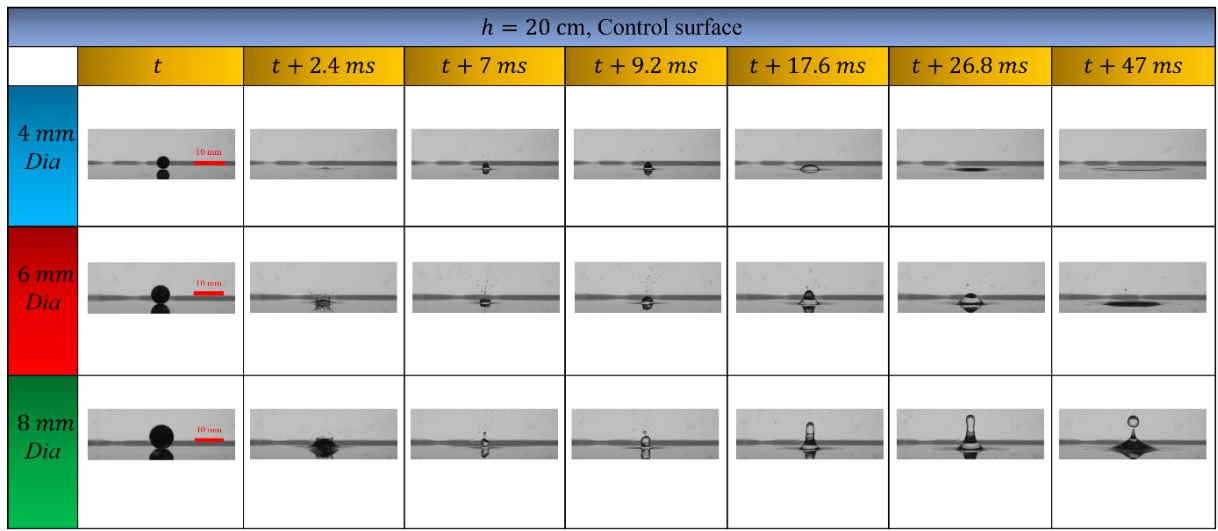

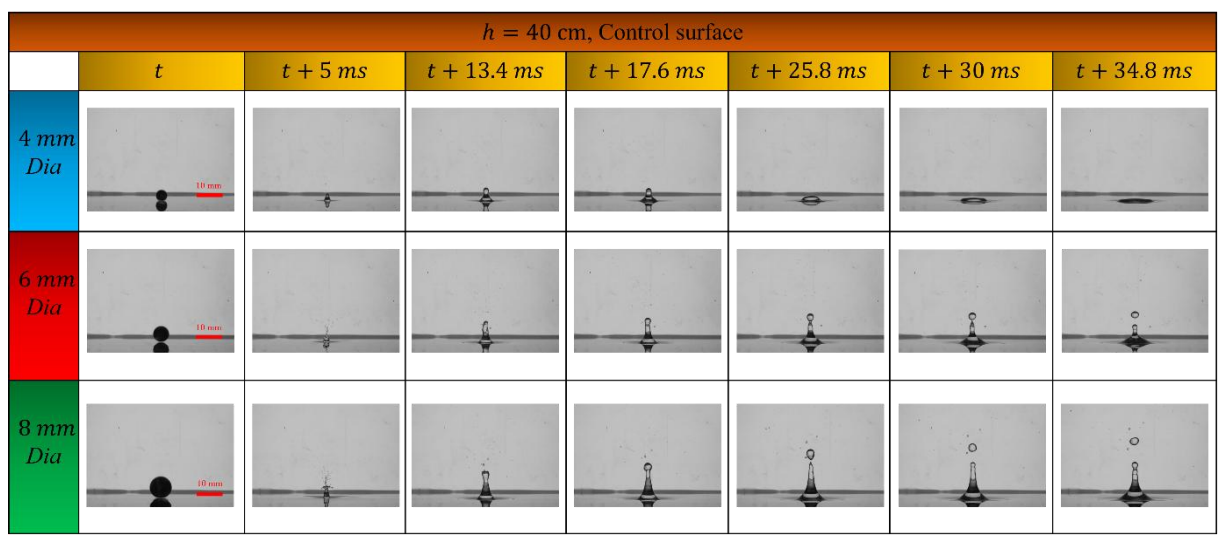

**Image S2.A. High-speed imaging (under water) of various diameter control surfaces with functionalization at different heights.**

The series of images depicts the time-resolved impact dynamics of control surfaces with OTS (octadecyltrichlorosilane) from different heights (5, 20, and 40 cm) and for varying spherical ball diameters (4, 6, and 8 mm) on water. On the control surface, the impact behavior shows relatively symmetric spreading, followed by retraction and rebounding, more prominently observed at lower heights and smaller droplet sizes. As impact height and droplet diameter increase, splashing and cavity formation become more pronounced. In contrast, the control with OTS surface demonstrates a dramatically different morphology. Owing to the hydrophobic nature of the OTS layer, the balls exhibit shallow cavity and the formation of Worthington jets, with minimal lateral spread. At higher heights, especially with larger diameters, deep seal cavity collapse and secondary droplet ejection are evident, highlighting enhanced air entrainment and energy transfer mechanisms. This comparative visualization emphasizes how both surface wettability and droplet kinetic energy play critical roles in dictating the impact outcome.

| | $t$ | $t + 7.8\ ms$ | $t + 9.2\ ms$ | $t + 9.6\ ms$ | $t + 13.4\ ms$ | $t + 15.6\ ms$ | $t + 18.2\ ms$ |
|---|---|---|---|---|---|---|---|
| 4 mm Dia | 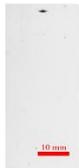 | 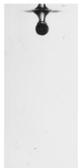 | 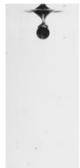 | 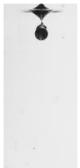 | 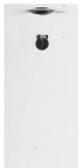 | 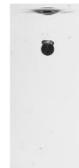 | 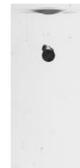 |
| 6 mm Dia | 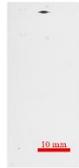 | 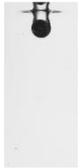 | 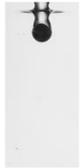 | 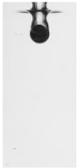 | 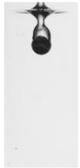 | 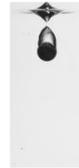 | 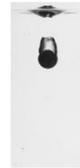 |
| 8 mm Dia | 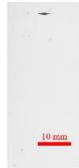 | 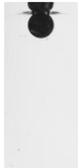 | 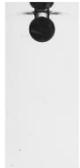 | 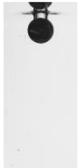 | 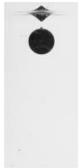 | 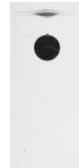 | 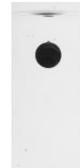 |

$h = 5$ cm, Control + OTS surface

| | $t$ | $t + 3.6\ ms$ | $t + 5.4\ ms$ | $t + 8.8\ ms$ | $t + 11.6\ ms$ | $t + 17.2\ ms$ | $t + 23.2\ ms$ |
|---|---|---|---|---|---|---|---|
| 4 mm Dia | 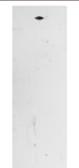 | 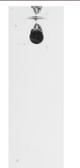 | 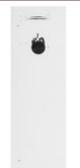 | 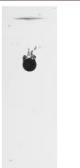 | 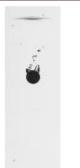 | 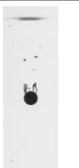 | 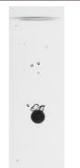 |
| 6 mm Dia | 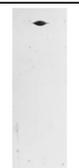 | 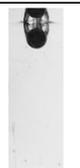 | 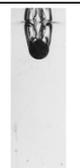 | 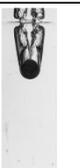 | 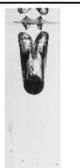 | 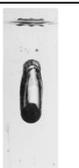 | 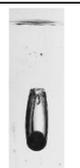 |
| 8 mm Dia | 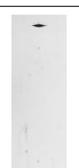 | 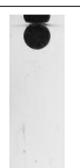 | 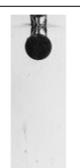 | 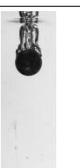 | 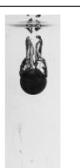 | 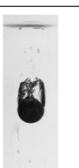 | 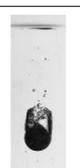 |

$h = 20$ cm, Control + OTS surface

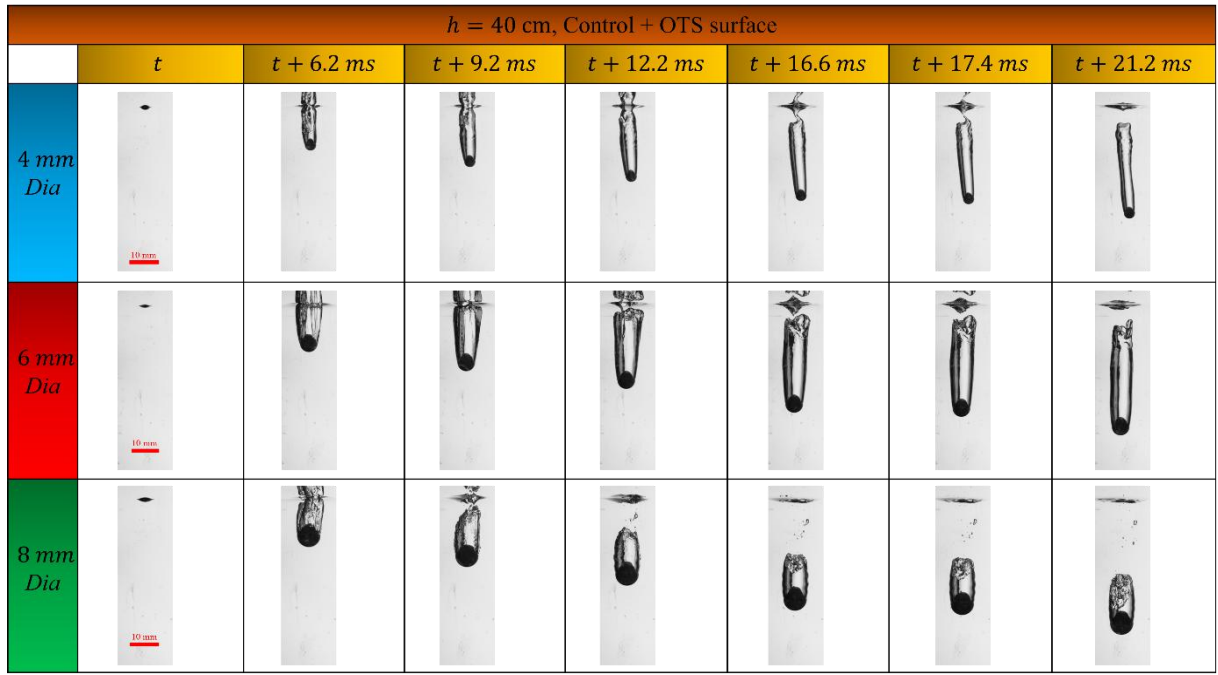

**Image S2.B. High-speed imaging (above the water surface) of various diameter control surfaces with functionalization at different heights.**

Similar dynamic behaviors were also observed above the water surface following the impact. In each case, the jet formation and spreading upon contact closely matched the patterns seen below the surface. For the 4 mm and 6 mm diameter droplets, minimal splash, with the interface remaining relatively stable. However, in the case of the 8 mm diameter, a Worthington jet was clearly visible above the surface, forming approximately 20.2 ms after impact and reaching its peak at around 25.8 ms. This jet formation mirrors the subsurface cavity collapse dynamics and is strongly influenced by ball size and impact inertia.

| | $t$ | $t + 3\ ms$ | $t + 8\ ms$ | $t + 14.4\ ms$ | $t + 17.4\ ms$ | $t + 20.2\ ms$ | $t + 25.8\ ms$ |
|---|---|---|---|---|---|---|---|
| | \multicolumn{7}{c|}{$h = 5$ cm, Control + OTS surface} |
| 4 mm Dia | 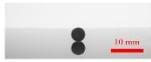 | 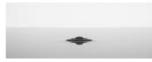 | 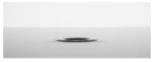 | 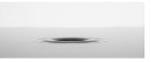 | 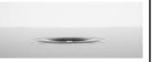 | 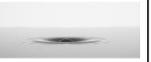 | 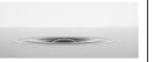 |
| 6 mm Dia | 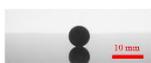 | 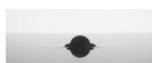 | 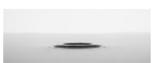 | 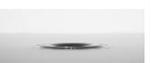 | 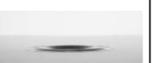 | 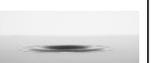 | 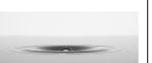 |
| 8 mm Dia | 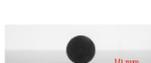 | 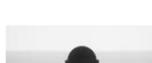 | 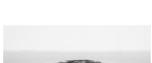 | 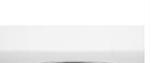 | 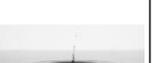 | 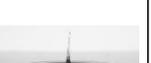 | 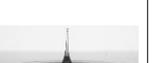 |

| | $t$ | $t + 1.6\ ms$ | $t + 2.8\ ms$ | $t + 6.2\ ms$ | $t + 13.6\ ms$ | $t + 19\ ms$ | $t + 32\ ms$ |
|---|---|---|---|---|---|---|---|
| | \multicolumn{7}{c|}{$h = 20$ cm, Control + OTS surface} |
| 4 mm Dia | 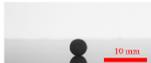 | 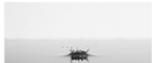 | 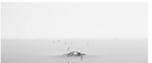 | 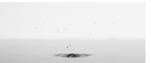 | 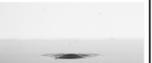 | 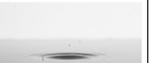 | 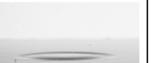 |
| 6 mm Dia | 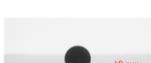 | 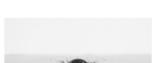 | 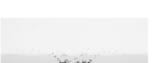 | 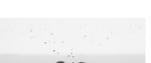 | 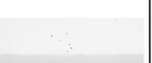 | 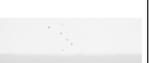 | 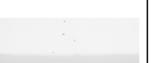 |
| 8 mm Dia | 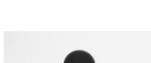 | 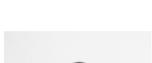 | 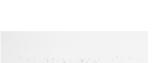 | 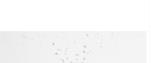 | 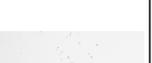 | 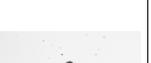 | 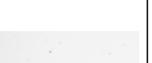 |

| | $t$ | $t + 1.2\ ms$ | $t + 2\ ms$ | $t + 5.4\ ms$ | $t + 8\ ms$ | $t + 24\ ms$ | $t + 34\ ms$ |
|---|---|---|---|---|---|---|---|
| | \multicolumn{7}{c|}{$h = 40$ cm, Control + OTS surface} |
| 4 mm Dia | 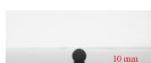 | 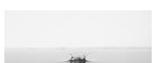 | 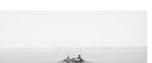 | 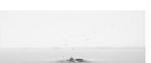 | 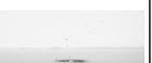 | 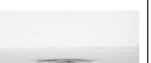 | 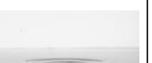 |
| 6 mm Dia | 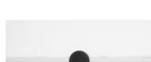 | 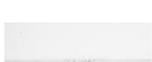 | 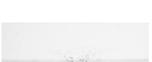 | 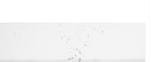 | 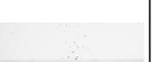 | 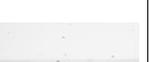 | 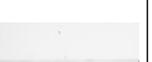 |
| 8 mm Dia | 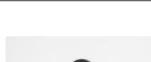 | 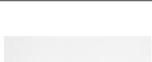 | 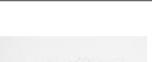 | 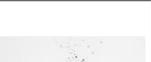 | 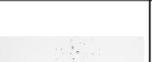 | 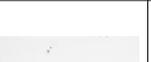 | 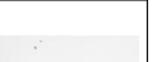 |

**Image S3.A. High-speed imaging (under water) of various diameters functionalized with textured (Micro-nano-textured) at different heights.**

These high-speed image sequences clearly demonstrate how ball diameter plays a critical role in governing the dynamics of impact and subsequent fluid interaction on the micro-nano textured OTS surface. For the smallest diameter (4 mm), the cavity is termed quasi-static, with a 5 cm height, whereas for a 20cm height, it is a shallow seal cavity. The largest diameter (8 mm) shows the most pronounced cavity at 40cm height and significant jet behavior, highlighting the amplified inertial effects and fluid entrainment associated with increased size. These observations underscore the strong dependence of solid impact phenomena on ball size, even under identical impact heights and surface treatments.

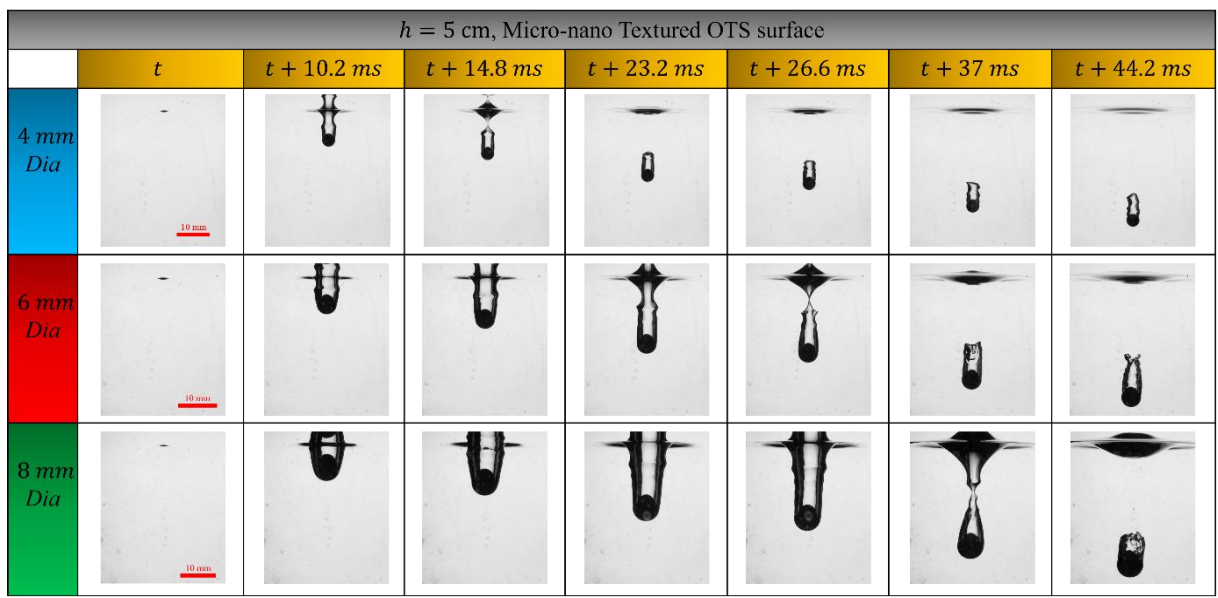

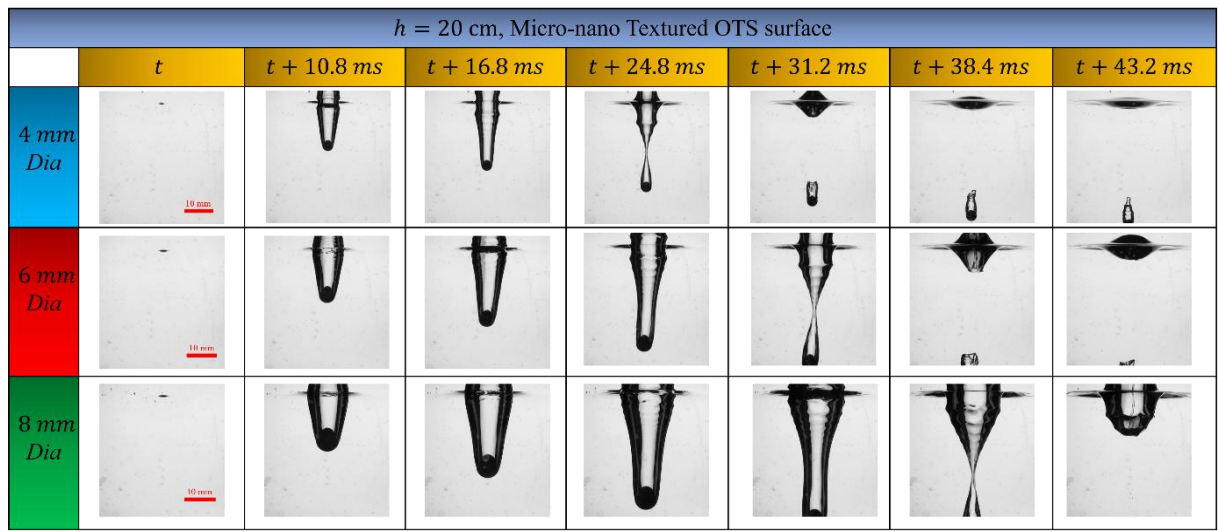

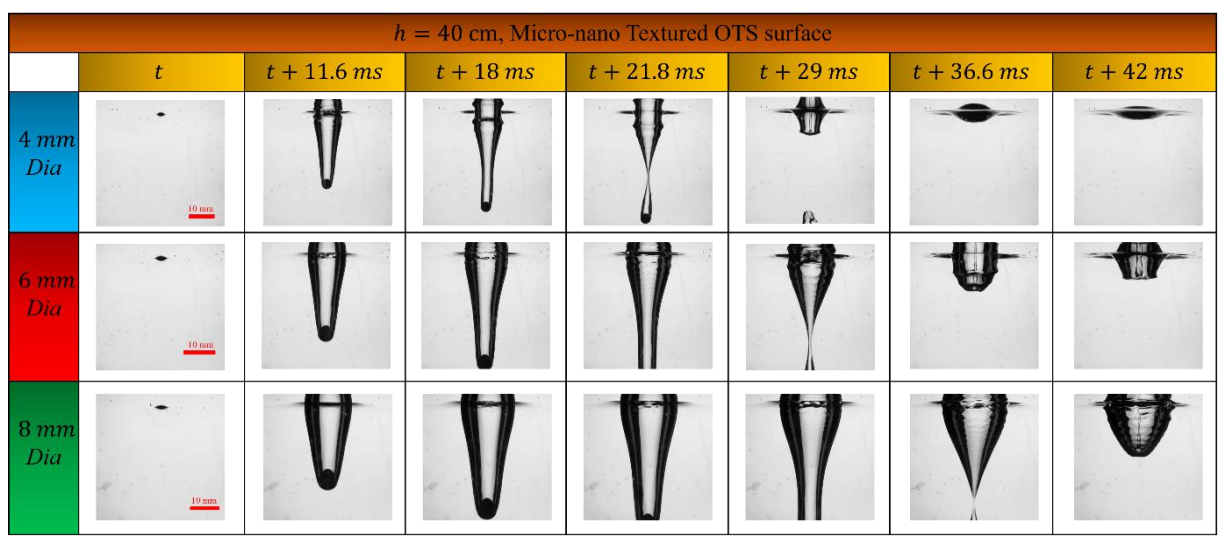

**Image S3.B. High-speed imaging (above the water surface) of various diameters functionalized with textured (Micro-nano-textured) at different heights.**

High-speed imaging (above the water surface) was employed to analyze the impact dynamics, such as the Worthington jet, on functionalized surfaces with varying diameters and micro-nano-textured topographies. This technique captured rapid interfacial interactions. By systematically altering texture scale, sphere diameter, and impact height, the study elucidated the role of hierarchical structures in modulating wettability and dynamic response.

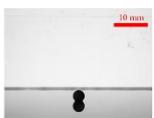

**Image S4.A. High-speed imaging (under water) of various diameter functionalized with textured (Nano-textured) at different heights.**

The image depicts a time-sequenced comparison of impact dynamics of a nano-textured OTS (octadecyltrichlorosilane) surface on water surface from various heights of 5, 20, and 40 cm, for spheres of three different diameters: 4, 6, and 8 mm. Each row represents a different diameter, while the columns show the progression of the solid sphere interaction with the water over time, from the initial moment of contact to a maximum of ~ 40 milliseconds post-impact. As observed, a 4 mm ball exhibits limited spreading, indicative of lower inertia. In contrast, larger diameters (6 and 8 mm) demonstrate more pronounced spreading and larger cavity formation, with the 8 mm one showing significant Worthington jet and splash-like behavior.

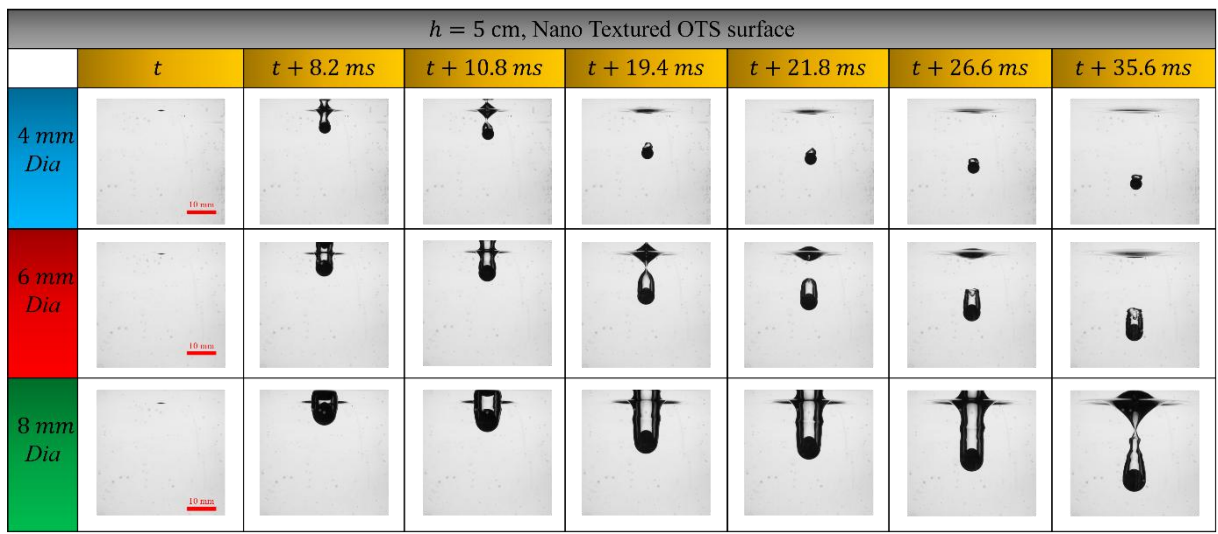

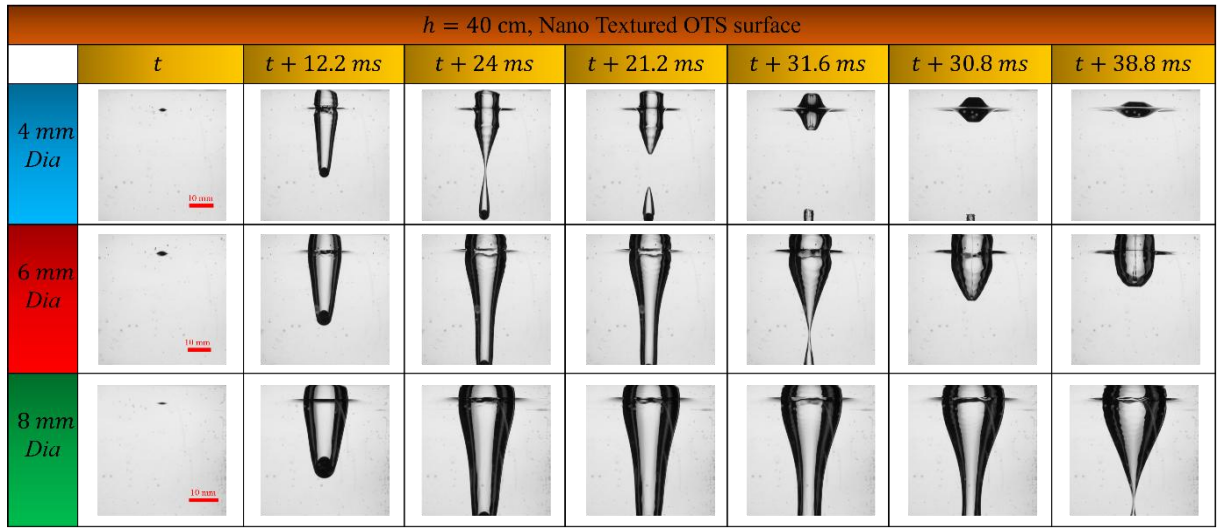

**Image S4.B. High-speed imaging (above the water surface) of various diameter functionalized with textured (Nano-textured) at different heights.**

The image sequence illustrates the side-view dynamics of solid impact of a nano-textured OTS surface from various heights of 5, 20, and 40 cm for three different sphere diameters: 4, 6, and 8 mm. The columns depict time-lapsed images from the moment of initial contact to ~ 70 milliseconds post-impact. The 4 mm spherical ball shows minimal spreading and rapid retraction, whereas 6 mm one demonstrates a more pronounced spreading phase followed by a noticeable Worthington jet phenomenon before collapsing. The 8 mm one exhibits significant spreading and flattening, forming a central upward jet. These observations emphasize the critical role of size and height in modulating energy dissipation, surface interaction, and cavity behavior on superhydrophobic nano-engineered surfaces.

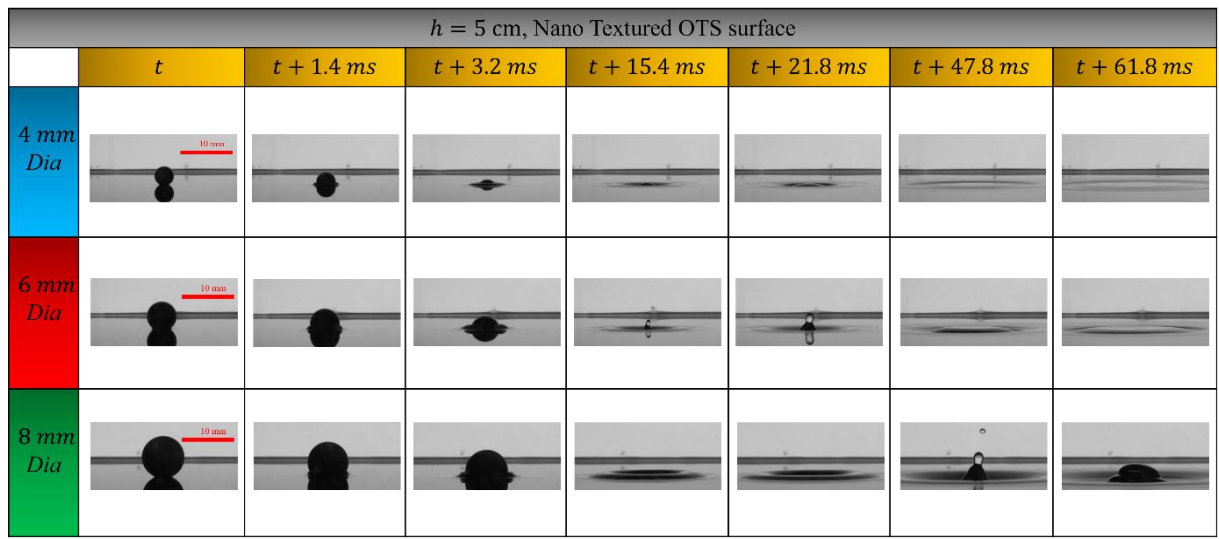

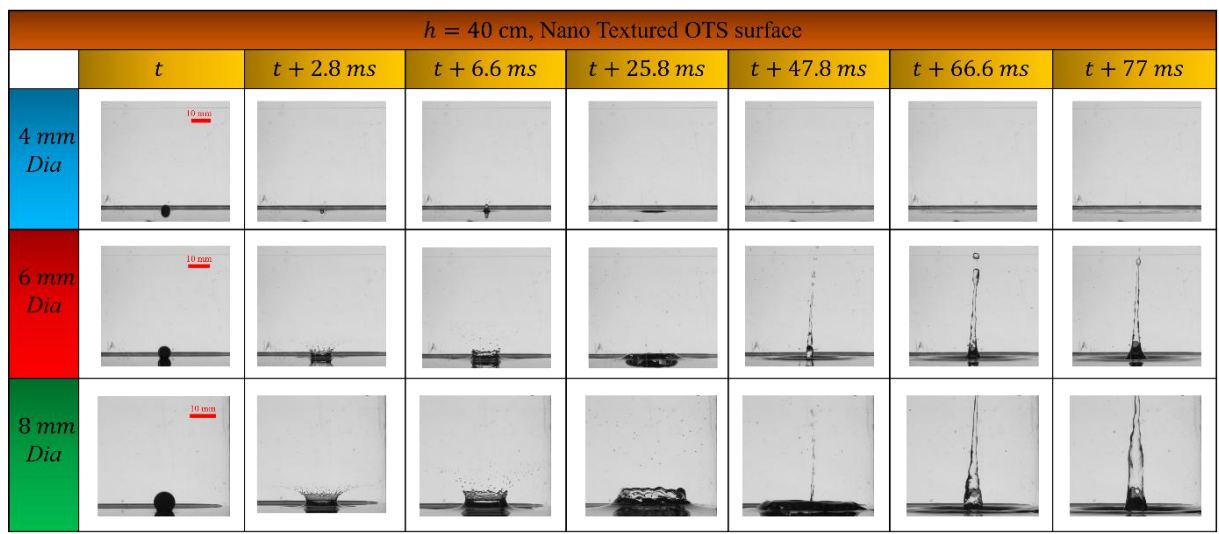

**Image S5.A. High-speed imaging (under water) of various diameters LIS (Micro-nano-textured) at different heights.**

The sequential high-speed images illustrate a micro-nano structured LIS impact on the water surface from various heights of 5, 20, and 40 cm with diameters of 4, 6, and 8 mm. The time-lapsed frames, captured from the moment of initial contact up to various maximum time scales such as 21.2, 38.4, and 42.6 milliseconds post-impact, reveal distinct cavity dynamic behaviors governed by height. Notably, larger balls demonstrate increased wetted area and prolonged interaction with the surface, indicating a stronger inertial influence compared to the smaller droplets. The interfacial energy between infused oil and water is lower than that of air and water. These observations underscore the critical role of size in modulating the interfacial dynamics on LIS surfaces, which is vital for underwater applications.

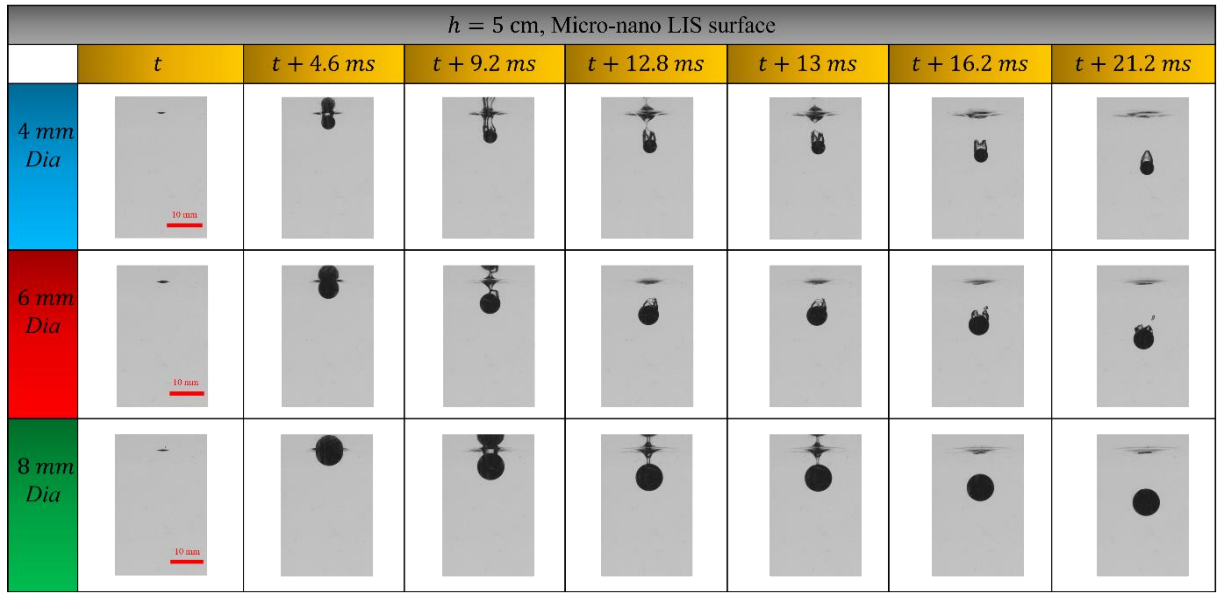

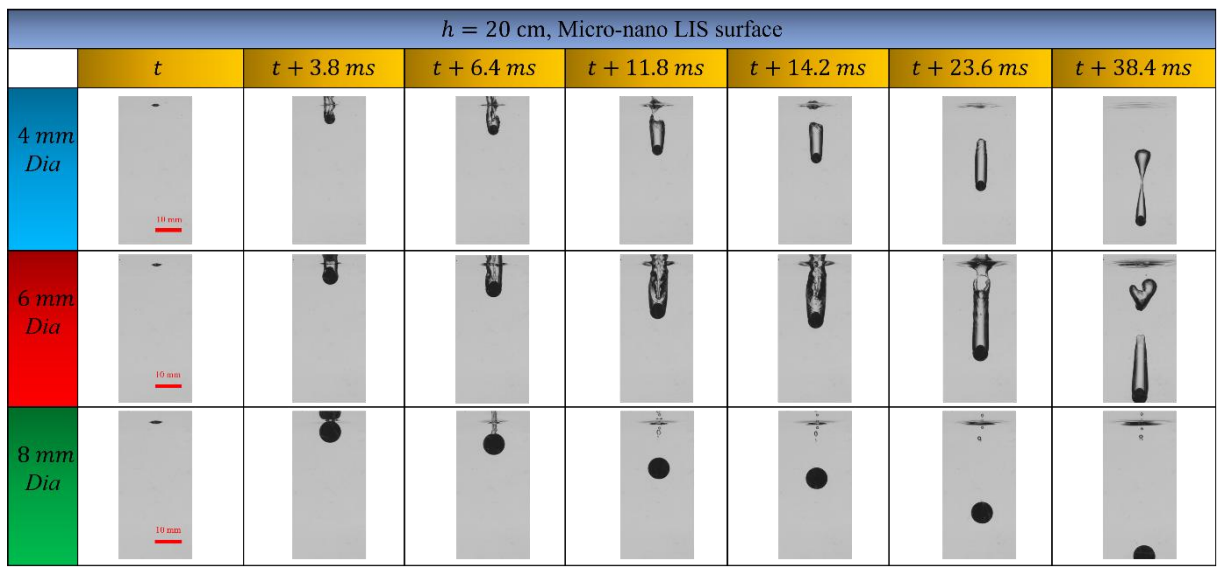

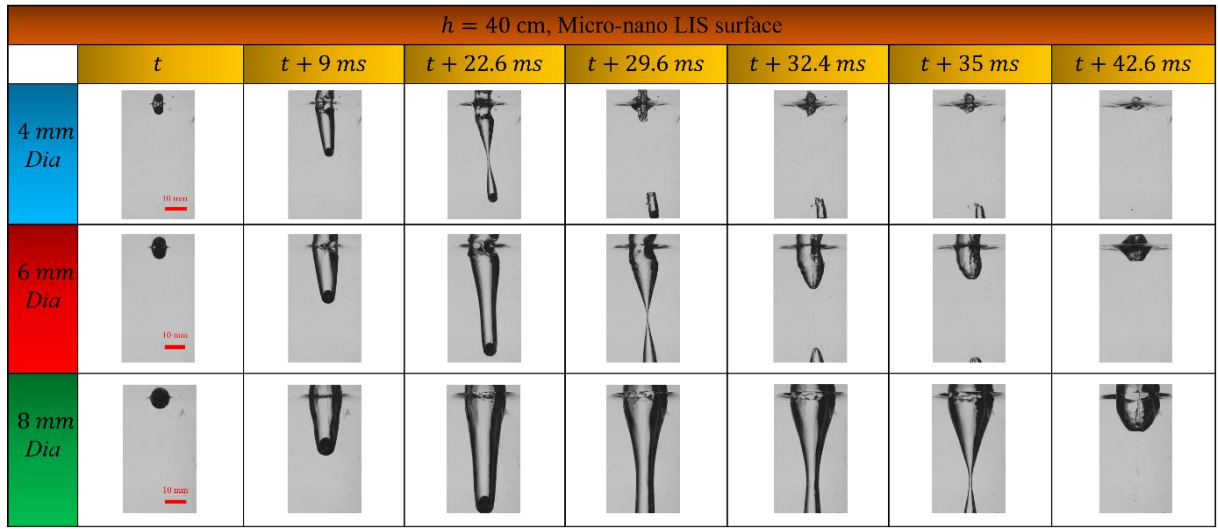

**Image S5.B. High-speed imaging (above the water surface) of various diameters LIS (Micro-nano-textured) at different heights.**

Noticeable Worthington jet on micro-nano textured LIS was observed from a height of 40 cm. The sequence highlights the spreading, cavity formation influenced by size and height. All balls exhibit a symmetric spreading phase followed by a central upward jet formation and subsequent collapsing. The 8 mm ball, in particular, shows a significant vertical jet and delayed collapse compared to the smaller ones. These results emphasize the scale-dependent impact dynamics of LIS, which are crucial for optimizing surface designs.

| | $t$ | $t + 3.4\ ms$ | $t + 10.2\ ms$ | $t + 12.8\ ms$ | $t + 20.8\ ms$ | $t + 35\ ms$ | $t + 47.8\ ms$ |
|---|---|---|---|---|---|---|---|
| | | | $h = 5$ cm, Micro-nano LIS surface | | | | |
| 4 mm Dia | 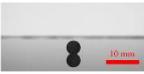 | 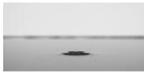 | 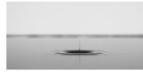 | 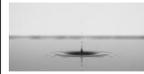 | 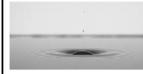 | 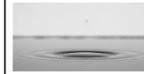 | 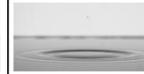 |
| 6 mm Dia | 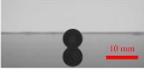 | 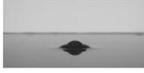 | 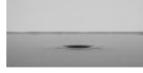 | 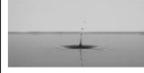 | 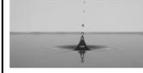 | 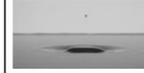 | 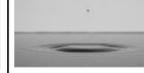 |
| 8 mm Dia | 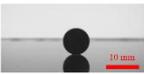 | 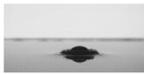 | 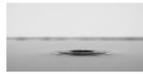 | 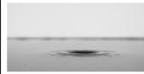 | 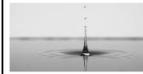 | 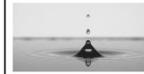 | 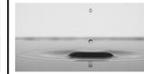 |

| | $t$ | $t + 3.4\ ms$ | $t + 6.2\ ms$ | $t + 8.8\ ms$ | $t + 15.4\ ms$ | $t + 18.8\ ms$ | $t + 31.8\ ms$ |
|---|---|---|---|---|---|---|---|
| | | | $h = 20$ cm, Micro-nano LIS surface | | | | |
| 4 mm Dia | 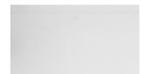 | 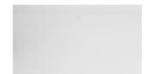 | 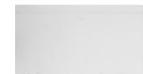 | 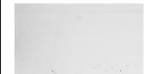 | 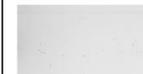 | 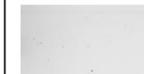 | 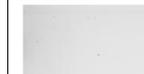 |
| 6 mm Dia | 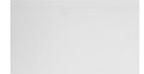 | 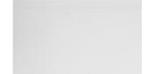 | 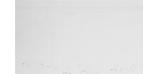 | 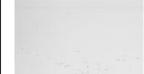 | 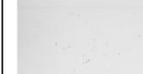 | 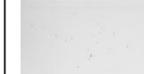 | 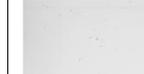 |
| 8 mm Dia | 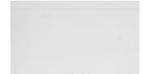 | 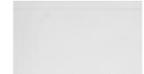 | 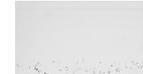 | 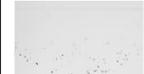 | 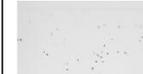 | 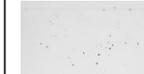 | 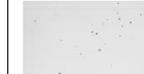 |

| | $t$ | $t + 3.4\ ms$ | $t + 9\ ms$ | $t + 17.6\ ms$ | $t + 33.4\ ms$ | $t + 39.6\ ms$ | $t + 48\ ms$ |
|---|---|---|---|---|---|---|---|
| | | | $h = 40$ cm, Micro-nano LIS surface | | | | |
| 4 mm Dia | 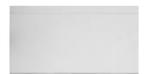 | 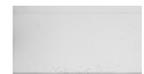 | 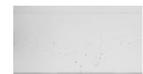 | 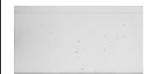 | 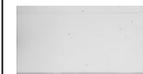 | 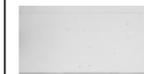 | 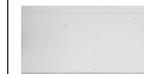 |
| 6 mm Dia | 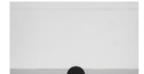 | 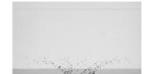 | 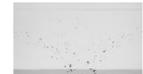 | 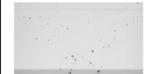 | 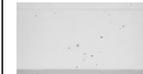 | 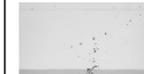 | 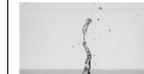 |
| 8 mm Dia | 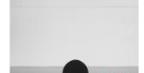 | 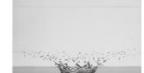 | 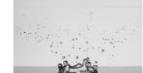 | 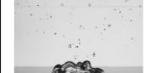 | 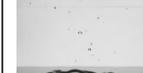 | 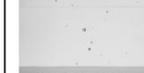 | 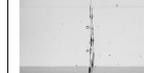 |

**Image S6.A. High-speed imaging (under water) of various diameter LIS (Nano-textured) at different heights.**

All diameter spherical balls exhibit extreme elongation; however, these are shallow cavities, vertical stretching, and the formation of well-defined cavities, followed by oscillatory retraction. Notably, 20 and 40 cm with an 8 mm ball diameter demonstrate larger cavity collapse and retraction dynamics, suggesting a transition to splash-dominated behavior at higher inertial regimes.

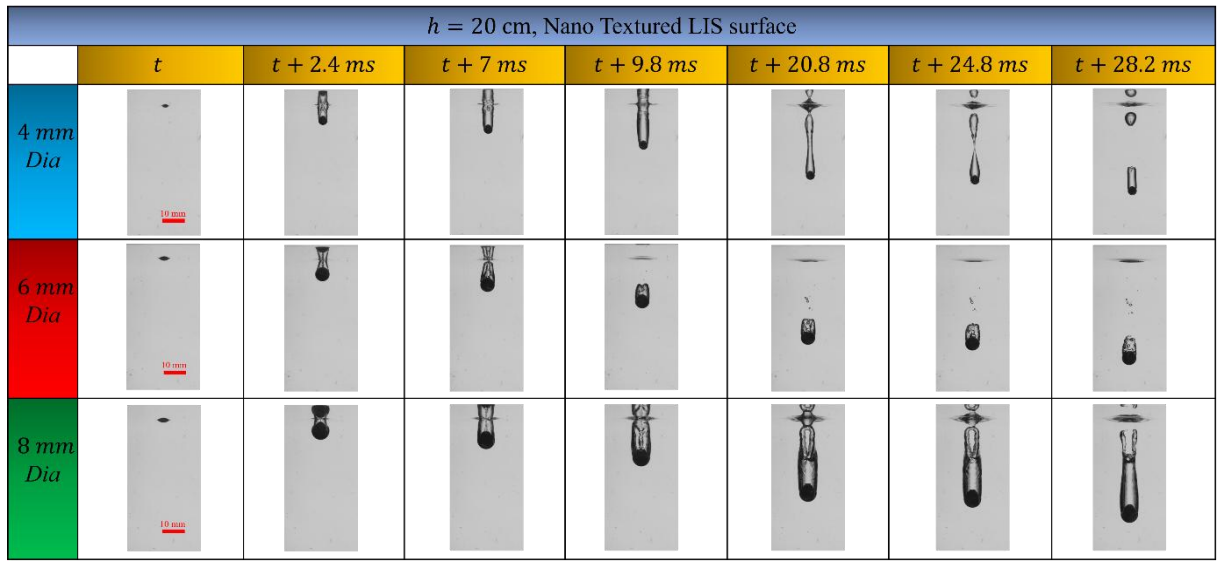

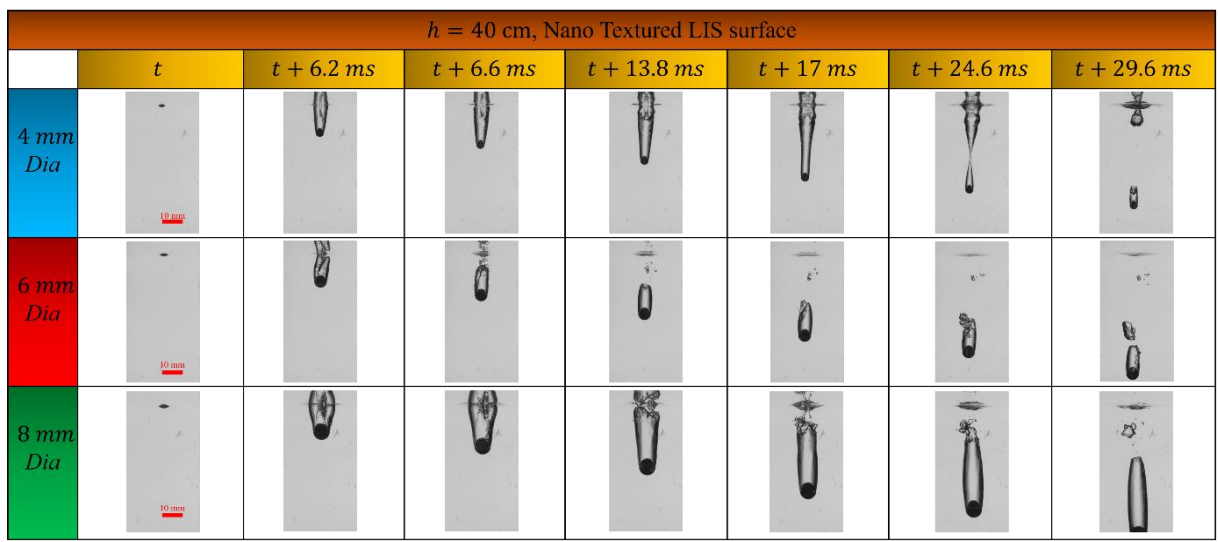

**Image S6.B. High-speed imaging (above the water surface) of various diameter LIS (Nano-textured) at different heights.**

The time-resolved sequence spans from the moment of contact, approximately 40 milliseconds post-impact, revealing the dynamic interaction between solid inertia and the surface characteristics. As the ball size increases, the splashing intensity and crown formation become more prominent, and the width of the jet is larger. The damped oscillation and smooth collapsing of the jet at the interface further highlight the role of the LIS in minimizing pinning and enhancing mobility.

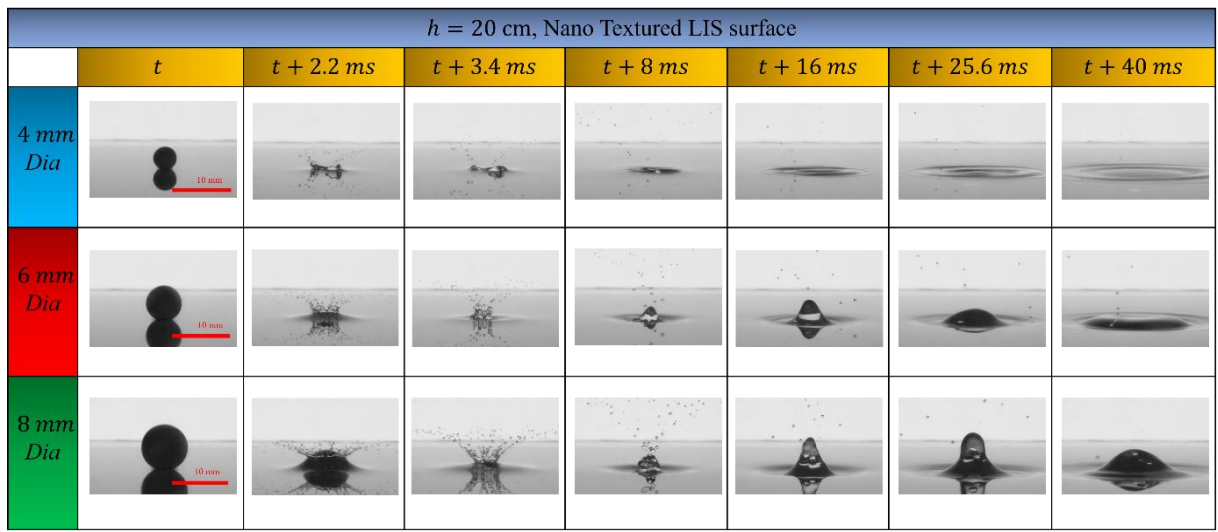

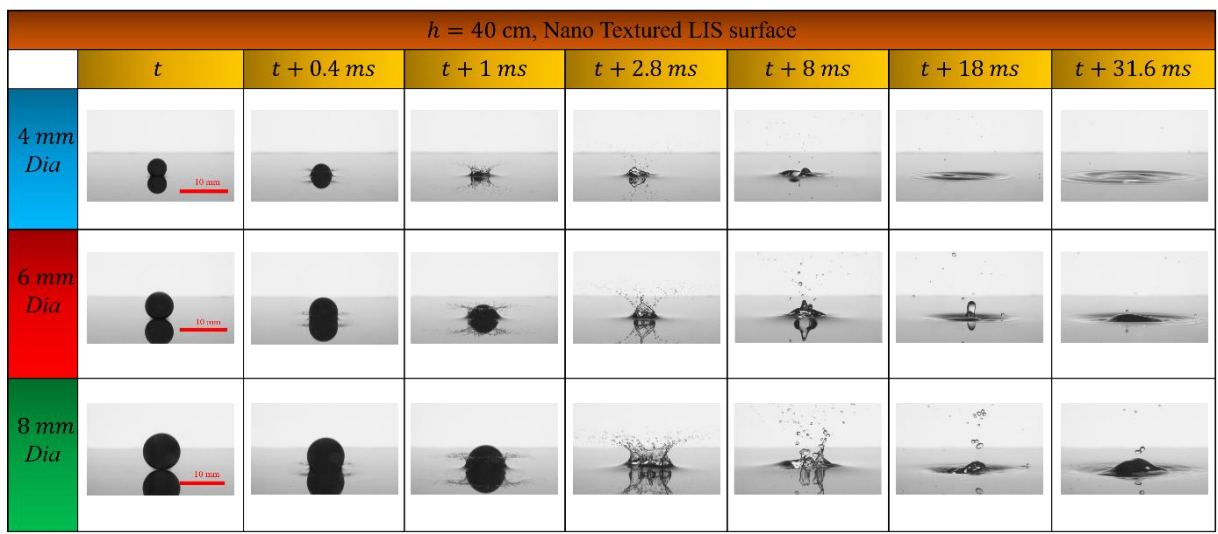